\documentclass[11pt,a4paper]{article}
 \usepackage{jheppub}
 \usepackage{graphicx}
\usepackage{amsmath,amssymb}
\usepackage{epsfig}

\def\x'{\mathaccent 19 x}
\def\y'{\mathaccent 19 y}
\def\n'{\mathaccent 19 n}
\def\u'{\mathaccent 19 u}

\def\et'{\mathaccent 19 \eta}
\def\th'{\mathaccent 19 \theta}
\def\lam'{\mathaccent 19 \lambda}
\def\varet'{\mathaccent 19 \vartheta}
\def\rh'{\mathaccent 19 \rho}
\def\ph'{\mathaccent 19 \Phi}
\def\xb'{\mathaccent 19 {\bar{x}}}

\def\det{\hbox{det}}
\def\be{\begin{equation}}
\def\ee{\end{equation}}

\newcommand{\bea}{\begin{eqnarray}}
\newcommand{\eea}{\end{eqnarray}}

\def\Tr{\text{Tr}}
\def\ha{\hat{a}}
\def\hm{\hat{m}}
\def\hz{\hat{z}}

\newcommand{\nn}{\nonumber}

\author{Gernot Akemann and}
\author{Fabrizio Pucci}
\affiliation{Fakult\"{a}t f\"{u}r Physik, Universit\"{a}t Bielefeld,
Universit\"{a}tsstra{\ss}e 25, D-33615 Bielefeld, Germany}
\small
\emailAdd{akemann@physik.uni-bielefeld.de}
\emailAdd{pucci@physik.uni-bielefeld.de}
\normalsize
\abstract{
We compute next-to-leading order (NLO) corrections in the $\epsilon$-regime of Wilson (WChPT) and Staggered Chiral Perturbation Theory (SChPT). A difference between the two is that in WChPT already at NLO, that is  at ${\cal O}(\epsilon^2)$, new low energy constants (LECs) contribute, whereas in SChPT they only enter at ${\cal O}(\epsilon^4)$.
We first determine the NLO corrections
in WChPT for $SU(2)$, and for $U(N_f)$ at fixed index.
This implies
corrections to the phase boundary  between the Aoki phase and the Sharpe-Singleton scenario in the thermodynamical limit
via corrections to the mean field potential.
We also compute
NLO corrections to the two-point function in the scalar and pseudo-scalar sector in WChPT.
Turning to SChPT we determine the NLO corrections to the LECs and their effect on the taste splitting. Here the NLO partition function can be written as the leading order one with renormalized couplings, thus preserving the equivalence to staggered chiral random matrix theory at NLO for any number of flavors $N_f$. In WChPT this relation only appears to hold for $SU(2)$.
}

\title{Exploring the Aoki regime}
\keywords{Wilson and staggered chiral perturbation theory, epsilon-regime, next-to-leading order effects}
\begin{document}
\maketitle
\renewcommand{\thefootnote}{\arabic{footnote}}
\setcounter{footnote}{0}

\section{Introduction}\label{intro}
In the study of the low energy dynamics of QCD it has
become
of considerable interest to extend chiral perturbation theory by including effects of the lattice spacing $a$.
Indeed in this way one can control how the ultraviolet cut-off influences every numerical simulations of lattice QCD. For what concerns the Wilson fermion formulation, in a series of works  \cite{Sharpe:1998xm,Rupak:2002sm,Bar:2003mh,Aoki:2004ta,Aoki:2003yv} it was shown how the continuum chiral Lagrangian gets modified from discretization effects, so-called Wilson Chiral Perturbation Theory (WChPT). For the staggered version of QCD the analogous extension has been done by Lee and Sharpe \cite{LeeSharpe} for one staggered flavor and then generalized by Aubin and Bernard
\cite{AubinBernard}, so-called Staggered Chiral Perturbation Theory (SChPT).
\newline
Quite recently there has been a lot of activity in the study of the  $\epsilon$-regime of WChPT and SChPT. The discretization effects compete with the quark mass for the explicit breaking of chiral symmetry. Despite the fact that this regime originally introduced in the continuum in \cite{GL87a} is unphysical, which is due to the fact that Compton wave-length is larger than the size of the box, it can be extremely useful in the determination of the low energy effective constants (LECs) and in the study of the Dirac operator spectrum and its dependence on the topology of the gauge fields \cite{Sharpe:2006ia,Damgaard:2010cz,Akemann:2010em,Splittorff:2011bj,Akemann:2011kj,Kieburg:2011uf,Akemann:2012pn,
Kieburg:2012yq}.
Due to the domination of the zero-mode group integral many computations can be performed analytically. First lattice simulations have already confirmed the quenched predictions \cite{Damgaard:2011eg,Deuzeman:2011dh,Damgaard:2012gy}.
\newline
Another intriguing characteristic of this regime is that at leading order (LO) in the $\epsilon$-expansion it is equivalent to a chiral Random Matrix Theory (ChRMT). More precisely, in the continuum theory a proof of this equivalence has been given in \cite{Damgaard:1998xy,Basile:2007ki} for all Dirac operator eigenvalue correlation functions.
The extension to theories at finite lattice spacing have been derived subsequently: in \cite{Damgaard:2010cz,Akemann:2010em,Splittorff:2011bj,Akemann:2011kj,Kieburg:2011uf,Akemann:2012pn,
Kieburg:2012yq} the authors showed that an analogous equivalence holds between the LO of WChPT and a Wilson Chiral Random Matrix Theory (WChRMT). In \cite{Osborn:2010eq}
this equivalence between the zero-mode sector of SChPT and the corresponding Staggered Chiral Random Matrix Theory (SChRMT) was established.
\newline
From the point of view of ChPT when introducing the effects of finite lattice spacing one has to consider the symmetry breaking
of the continuum theory down to a subgroup. Consequently more operators will be allowed in the Symanzik effective action, and more terms appear in the chiral Lagrangian.
In addition to the chiral condensate $\Sigma$ and to the
pion decay constant $F$
some new LECs
need to be introduced to characterize the strength of these terms.
Using the following power-counting in terms of the finite volume $V\sim \epsilon^{-4}$, $m\sim \epsilon^4,\, a\sim \epsilon^2$,
in WChPT in general 3 new LECs are introduced at LO and labeled by $W_6, W_7, W_8$, whereas for SChPT 6 new LECs need to be introduced, $C_1, C_2, C_3, C_4, C_5, C_6$.
The absolute sign of individual LECs and of their combinations has been subject of intense recent discussions in WChPT \cite{Akemann:2010em,Hansen:2011kk,Hansen:2011mc,Kieburg:2012fw}. It remains to be seen if such arguments carry over to NLO.
\newline
In the $\epsilon$-regime there are different ways of introducing the relative strength between the mass terms and the lattice spacing. In this paper we will use the so-called Aoki or large cut-off effect (LCE) counting where $m \sim a^2 \Lambda_{QCD}^3$ or $a\sim {\cal O}(\epsilon^2)$ \cite{Aoki:2007es}.
Thus the discretization terms are of the same order as the mass terms and compete for the breaking of chiral symmetry already at LO. Considering the Aoki regime means that the LO integral over the zero-modes is modified with respect to the continuum theory and makes the analytic computations more involved. However, one could also consider the Generic Small Mass (GSM) counting
\cite{Sharpe:2004ps,Sharpe:2004ny}
where the discretization errors enter only at next-to-next-to-leading order (NNLO) with respect to the continuum Lagrangian, since the counting is $m \sim a\Lambda_{QCD}^2$ or $a\sim {\cal O}(\epsilon^4)$. Finally an intermediate regime is known as the GSM* counting, where
$a\sim {\cal O}(\epsilon^3)$ and the discretization errors are at next-to-leading order (NLO) with  respect to the continuum \cite{Bar:2008th,Shindler:2009ri}.
\newline
The main question addressed in this paper is to extend WChPT and SChPT to order ${\cal O}(\epsilon^2)$, for both the partition functions, and for the scalar and pseudoscalar current correlators in the Wilson case.
For the spectral density of the Dirac operator a NLO order calculation has already been done, however not in the $\epsilon$- but in the $p$-regime \cite{Necco:2011vx}.
In contrast to the $\epsilon$-regime in the
continuum, sectors of fixed topological charge $\nu$ are no longer well defined at finite-lattice spacing. They have to be replaced by the index of the Dirac operator, and we refer to \cite{Akemann:2010em} for a detailed discussion and references.
In the continuum it was found that the NLO partition function could be written as the LO order one with renormalized couplings that include the order ${\cal O}(\epsilon^2)$ effects, both for $\Sigma$
and for $F$ when including a iso-spin chemical potential \cite{DGF,Akemann:2008vp,CT1}.
This implied that the partition function from ChPT and from ChRMT agree up to NLO. Only at NNLO non-universal effects were found in \cite{CT2}.
We will find that for WChPT only for $SU(2)$ the NLO contributions are absorbed into the two effective couplings relevant in that case,
whereas for $U(N_f)$ at fixed index we have to take extra derivatives of the LO partition function.
In contrast for SChPT the NLO contributions can be absorbed into renormalized couplings for any $N_f$, due to the $U(1)$ remnant
of the continuum chiral symmetry.
\newline
Other information can be extracted from the NLO partition function. For $SU(2)$ one can see that the
corrections can drive the system in or out of the Aoki phase, compared to LO. Depending on the coefficients of the mean field potential the theory  can stay in the Aoki phase where two pions are massless as a consequence of the breaking of the flavor symmetry, or in the Sharpe-Singleton scenario where a first order transition is present. The
NLO corrections can modify  the boundary of these two regions in the thermodynamical limit.
\newline
For what concerns staggered fermions the effective LECs we compute can lead to the following prediction. Since from the tree level Lagrangian one can see how the taste symmetry is broken, the new renormalized LECs allow to  quantify  how the finite-volume corrections can modify the taste symmetry violation.
\newline
The outline of the paper is as follows. In section \ref{WChPT}
we study WChPT at NLO starting with $SU(2)$ and including the effect on the phase boundary in subsection \ref{WChPTNf2}, and then turn to $U(N_f)$ in \ref{nu}. The two-point functions are given in \ref{WChPT2pt}. In section \ref{SChPT} we repeat our analysis for the staggered version, which includes the effect on the taste splittings.
Finally in the last section \ref{summary} our discussion and some considerations regarding possible extensions of this work are presented.
Several technical details are deferred to the appendices \ref{Agroup-id} to \ref{D-SChPT}.

\section{Wilson Chiral Perturbation Theory at NLO}\label{WChPT}

\subsection{Introduction}\label{introW}
In this section we consider the $\epsilon$-regime of Wilson Chiral Perturbation theory (WChPT) with $N_f=2$ degenerate quarks of mass $m$.
As already pointed out in the previous section and shown in \cite{Akemann:2010em} at LO it is equivalent to a Random Matrix Theory
which includes order $\mathcal{O}(a^2)$ discretization effects (WChRMT). Our aim is to analyze WChPT at the next-to-leading order (NLO) in the $\epsilon$-expansion and show that within the Aoki regime the partition function at that order can be rewritten as the LO one with renormalized low energy constants (LECs). In the continuum a similar relation between the LO and NLO partition function holds for every number of flavors as shown e.g. in \cite{DGF,Akemann:2008vp}.\newline
Let us start by introducing the two-flavor Wilson chiral Lagrangian that at LO in the Aoki regime can be written as
\begin{equation}
  \mathcal{L}_{\rm LO} =\frac{F^2}{4} \text{Tr}\left[ \partial_{\mu} U \partial_{\mu} U^\dag \right] - \frac{\Sigma}{2}  \text{Tr}\left[ M^{\dagger} U + U^{\dagger} M \right]
  + a^2 c_2\,\left(\text{Tr}\left[ U + U^{\dagger}\right]\right)^2 \ .\label{actionnf2}
\end{equation}
\noindent
In addition to the continuum Gasser-Leutwyler terms \cite{Gasser:1984gg,Weinberg:1978kz} there is an additional
order ${\cal O}(a^2)$ contribution and thus a new low energy effective constant $c_2$ (note that our
$c_2=W_6+W_8/2$ is a short hand notation for the standard terminology which is $c_2F^2/16$, and likewise for the other NLO LECs).
Here as usual $F$ is the pion decay constant, $\Sigma$ is the chiral condensate and $M$ is the mass matrix that, for two degenerate quarks with mass $m$, reduces to $M= m\, \mathbb{I}_{2 \times 2}$.
In all the following we will only consider degenerate masses.
We use the standard parameterization for the Goldstone boson
\begin{equation}
U(x) = U_0\, \, \text{exp}\left[i \frac{\sqrt{2}}{F}\, \xi(x) \right]\ ,
\label{Upara}
\end{equation}
\noindent
where $U_0$ is the two by two unitary matrix describing the zero-modes nonperturbatively, and the Hermitian fields $\xi(x) = \xi(x)^{\dagger} = \sigma_b \xi_b$, that belong to the Lie algebra $su(2)$, parameterize the propagating modes. In order to derive the Lagrangian (\ref{actionnf2}) in the so-called Aoki regime one has to use the power counting \cite{Aoki:2007es}
\begin{equation}
V \sim \epsilon^{-4},\, m\sim \epsilon^4,\, \partial \sim \epsilon, \, \xi(x)\sim \epsilon,\, \, a\sim \epsilon^2. \label{powercounting}
\end{equation}
\noindent
Other counting schemes can be considered if we want to study the GSM$^*$ or GSM regime \cite{Sharpe:2004ny,Sharpe:2004ps} that are in fact defined considering the cut-off effects respectively as NLO and NNLO contributions with respect to the continuum terms.\newline
If we want to go further in the Aoki regime and compute the NLO partition function we have to consider in addition to (\ref{actionnf2}) the NLO chiral Lagrangian.

\begin{center}
\begin{tabular}{|c|c|}
  \hline
  Leading Order $\mathcal{O}(\epsilon^0)$ & $m$, $p^2$, $a^2$ \\
  Next-to-Leading Order $\mathcal{O}(\epsilon^2)$ & $a m$, $a p^2$, $a^3$ \\
  Next-to-Next-to-Leading Order $\mathcal{O}(\epsilon^4)$ & $m^2$, $m p^2$, $p^4$, $a^2 m$, $a^2 p^2$, $a^4$ \\
  \hline
\end{tabular}
\end{center}
\begin{center}
\small
\textbf{Table 1}. Contributions to the Wilson Chiral Lagrangian in the Aoki regime, see \cite{Hansen:2011mc} for an explicit list of terms.
\end{center}
\normalsize
\noindent
As we can see from the table 1. that schematically gives us the counting of all the terms that contribute up to $\mathcal{O}(\epsilon^4)$ to the chiral Lagrangian, at order ${\cal O}(\epsilon^2)$ the possible terms that enter are  $\mathcal{O}(a p^2)$, $\mathcal{O}(a m)$ and $\mathcal{O}(a^3)$.
Following \cite{Gasser:1983yg,Aoki:2008gy,Hansen:2011mc,Bar:2003mh} they can be written as
\begin{eqnarray}
  \mathcal{L}_{\rm NLO} &=&  a\, c_0\,
\text{Tr}\left[ \partial_{\mu} U\, \partial_{\mu} U^{\dagger}\right]
\text{Tr}\left[ U +  U^{\dagger}\right]
+ a m\, c_3\, \left(\text{Tr}\left[ U +  U^{\dagger}\right]\right)^2 \nn\\
&&+ a^3 d_1 \, \text{Tr}\left[ U +  U^{\dagger}\right] + a^3 d_2 \left(\text{Tr}\left[ U +  U^{\dagger}\right]\right)^3 \label{actionnf2NLO}
\end{eqnarray}
\noindent
where for $SU(2)$ 4 new and undetermined LECs, namely $c_0, c_3, d_1$ and $d_2$, need to be introduced. Here we have explicitly used some special properties of $SU(2)$, see e.g. appendix \ref{Agroup-id}, compared to the general $N_f$ case.
Now using the power counting (\ref{powercounting}) we expand the action
\begin{equation}S = \int d^4x\, \left( \mathcal{L}_{\rm LO} + \mathcal{L}_{\rm NLO} \right)\end{equation}
\noindent
up to $\mathcal{O}(\epsilon^2)$, where we  obtain
\begin{eqnarray}
S^{(0)} &=& \frac{1}{2}\, \int d^4x\, \text{Tr}\left[\partial_{\mu} \xi(x) \partial_{\mu} \xi(x)\right]
-  \frac{1}{2}\, m\, V \Sigma\, \text{Tr}\left[ U_0 + U_0^{\dagger} \right]
+ a^2 V c_2\, \left(\text{Tr}\left[ U_0 + U_0^{\dagger}\right]\right)^2\nn\\
&\equiv& S^{(0)}_{\partial^2}+S^{(0)}_{U_{0}}
\end{eqnarray}
\noindent
for the $\mathcal{O}(\epsilon^0)$ contribution. Here we
have defined the LO part of propagating and zero-modes separately.
For the $\mathcal{O}(\epsilon^2)$ terms we get
\begin{eqnarray}
S^{(2)}&=&  \frac{1}{12\, F^2} \int d^4x \text{Tr}\Big[[\partial_{\mu} \xi(x), \xi(x)][\partial_{\mu} \xi(x), \xi(x)]\Big]
+ \frac{m \Sigma}{2 F^2}\int d^4x\, \text{Tr}\left[ \left(U_0 + U_0^{\dagger}
\right) \xi(x)^2  \right]
\nn\\ 
&&- 2 a^2  \frac{ c_2 }{F^2} \int d^4x\, \left(\text{Tr}\left[ \left( U_0 - U_0^{\dagger}\right) \xi(x)\right]\right)^2\nn\\
&&- 2 a^2\frac{ c_2 }{F^2}\text{Tr}\left[ U_0 + U_0^{\dagger}\right]
 \int d^4x \text{Tr}\left[ \left(U_0 + U_0^{\dagger }\right) \xi(x)^2\right]\nn\\
&&+ \frac{2 a\, c_0}{F^2}
\text{Tr}\left[ U_0 +  U_0^{\dagger}\right]
\int d^4x\, \text{Tr}\left[ \partial_{\mu} \xi(x)\, \partial_{\mu} \xi(x) \right]
\ +\  a m\, c_3\, V\, \left(\text{Tr}\left[ U_0 +  U_0^{\dagger}\right]
\right)^2\nn\\
&&+ a^3 d_1 \, V\, \text{Tr}\left[ U_0 +  U_0^{\dagger}\right]
+ a^3 d_2\, V\,\left(\text{Tr}\left[ U_0 +  U_0^{\dagger}\right]\right)^3\ ,
\end{eqnarray}
\noindent
while the $\mathcal{O}(\epsilon)$ term  vanishes due to $\int d^4x\, \xi(x)=0$.

\subsection{Partition Function for $N_f=2$ and Aoki Phase
}\label{WChPTNf2}

The next step is the calculation of the partition function up to  $\mathcal{O}(\epsilon^2)$. The general form of the partition function can be rearranged by separating the integration over the zero-modes from the integration over the Gaussian fluctuations as
\begin{equation}\mathcal{Z} = \int_{SU(2)} [d_H U(x)]\, e^{-S} = \int_{SU(2)} d_H U_0\, \, e^{-S_{U_0}^{(0)}}\, \, \mathcal{Z}_{\xi}(U_0)\ ,
\end{equation}
\noindent
with
\begin{equation}\mathcal{Z}_{\xi}(U_0) = \int [d\xi(x)]\, \left( 1 - \frac{2}{3 F^2 V} \int d^4x\, \text{Tr}\left[\xi(x)^2\right]\right) e^{S_{U_0}^{(0)} - S},\end{equation}
\noindent
containing the Jacobian $J(\xi(x))$ up to order $\mathcal{O}(\epsilon^3)$
\cite{GL87a} from the parameterization (\ref{Upara}),
and the chiral action $S$ to an unspecified order. Here the invariant Haar measure
\begin{equation}[d_{H} U(x)] = d_{H} U_0\, [d\xi(x)]\, \left( 1 - \frac{N_f}{3 F^2 V} \int d^4x \text{Tr}\left[\xi(x)^2\right]\right)
\label{Jacobian}
\end{equation}
\noindent
has been divided as the invariant measure over the zero-modes $U_0$ times the flat measure over the
fluctuations $\xi(x)$. At this point one can expand
the function $Z_{\xi}(U_0)$ up to $\mathcal{O}(\epsilon^2)$
and then perform all the Gaussian integrals using the expression
\begin{equation}
\int [d\xi(x)]
\exp[-S_{\partial^2}^{(0)}]\
\xi(x)_{ij}\xi(y)_{kl} = \left( \delta_{il}\delta_{jk} - \frac{1}{N_f}\, \delta_{ij}\delta_{kl}\right)\Delta(x-y)\label{propagator}
\end{equation}
\noindent
in terms of the propagator.  We easily find that
\begin{eqnarray}
\nonumber
{\cal Z}_{\xi}(U_0) &=& \mathcal{N} \left\{ 1+\left(
-\, \frac{3\, m V \Sigma}{4 F^2} \Delta(0)
- a^3 d_1 V  \right)
 \text{Tr}\left[ U_0 + U_0^{\dagger} \right]
 \right. \\
&& \left.
+ \left(\frac{4 a^2 c_2 V}{ F^2}  \Delta(0)- a m c_3V \right)\left(\text{Tr}\left[ U_0 +  U_0^{\dagger}\right]\right)^2  -a^3 d_2 V\left(\text{Tr}\left[ U_0 +  U_0^{\dagger}\right]\right)^3\right\}.
\nn\\&&
\label{3211}\end{eqnarray}
\noindent
Here $\mathcal{N}$ is an overall normalization factor that contains all
constants that are $U_0$ independent and that drop out in expectations values.
In particular this includes the contribution from the Jacobian.
The propagator $\Delta(0)$ is finite in dimensional regularization and
is given by
$\Delta(0) = -\beta_1/\, V^{1/2}$, with $\beta_1$ a numerical coefficient that encodes the geometrical data of the box considered. At this point we note that all terms in (\ref{3211}) can be reabsorbed easily in the LO chiral Lagrangian by re-exponentiating the corrections, with the only
exception of the last term. In order to solve this problem one can write this contribution as a sum of single and double trace terms using the relation (\ref{SecondaSU2}) obtained in the appendix \ref{A-su2}
through some group integral identities. Finally the partition function can be written as
\bea
  \mathcal{Z}_{\rm NLO} &=&\mathcal{N}\ ' \int_{SU(2)} d_H U_0\, \text{exp}\left[  \frac{m \Sigma^{\text{eff}} V}{2} \text{Tr}\left[ U_0 + U_0^{\dagger} \right]
  - a^2\, c_{2}^{\text{eff}}\, V\left(\text{Tr}\left[ U_0 + U_0^{\dagger}\right]\right)^2\right]\ \ \nn\\
&=&\frac{\mathcal{N}\ ' }{\mathcal{N}}\
 \mathcal{Z}_{\rm LO}(\Sigma^{\text{eff}},c_{2}^{\text{eff}})\ ,
\label{ZNLOnf2}
\eea
\noindent
with the effective renormalized LECs given by
\begin{equation}\Sigma^{\text{eff}}
=\Sigma \left( 1
- \frac{3}{2 F^2}\Delta(0)-\frac{\ha}{\hm\sqrt{V}}
\left(2\ha^2d_1+32\ha^2d_2-3\frac{d_2}{c_2}\right)\right)
\ ,
\end{equation}
\noindent
and
\begin{equation}c_{2}^{\text{eff}}
= c_2 \left( 1 -   \frac{4}{ F^2}\Delta(0)\right)+\frac{\hm}{\ha}
\left(\frac{c_3}{\Sigma}+\frac{d_2}{4c_2}\right)\frac{1}{\sqrt{V}}\ .
\end{equation}
\noindent
Here we have defined
\be
\hm\equiv m\Sigma V\ \ \mbox{and} \ \ \ha^2\equiv a^2V\ ,
\label{V-scale}
\ee
which are of order ${\cal O}(1)$.
Differently from the continuum limit chiral perturbation theory, in which the NLO renormalized LECs can be rewritten only as functions of the number of flavors and the geometry of the system, here their expressions involve also some NLO LECs. In principle
this allows us to extract them from lattice computations through a finite-size scaling analysis. Performing the simulations at two different lattice volumes $V_1$ and $V_2$, with geometries
$\beta_1$ and $\beta_2$, WChPT predicts a scaling of the LECs as
\begin{eqnarray}\frac{\Sigma^{\text{eff}}(V_1)}{\Sigma^{\text{eff}}(V_2)}
&\approx& 1
+ \frac{3}{2 F^2} \frac{(\beta_1 \sqrt{V_2}- \beta_2 \sqrt{V_1}) }{\sqrt{V_1 \, V_2}}
+ \left(\frac{3 a d_2 }{m c_2 \Sigma}
\right)
\left( \frac{1}{V_1} - \frac{1}{V_2}\right)
+{\cal O}\left(\frac{1}{V}\right)\ ,
\\
\frac{c_{2}^{\text{eff}}(V_1)}{c_{2}^{\text{eff}}(V_2)} &\approx& 1 + \frac{4}{F^2} \frac{(\beta_1 \sqrt{V_2}- \beta_2 \sqrt{V_1} )}{\sqrt{V_1 \, V_2}}
+{\cal O}\left(\frac{1}{V}\right)
\ ,
\end{eqnarray}
\noindent
where we have given the ${\cal O}(1)$ and ${\cal O}(1/\sqrt{V})$ terms in the scaling limit eq. (\ref{V-scale}).
From the NLO partition function we can also extract information about the Aoki phase. The possible existence of such a phase in which flavor symmetry can be broken (with no analogon in the continuum theory) has been an outstanding problem for a long time. Quite recently in \cite{Kieburg:2012fw} the authors showed that while in the
 unquenched theory both scenarios (the Aoki and Sharpe-Singleton scenario) can be realized, the quenched theory at sufficiently small quark mass
 is always in the Aoki phase (see \cite{Bernardoni:2011fx,Aoki:1992nb,Aoki:1989rw,Aoki:1990ap,Jansen:2005cg,Aoki:1995yf,Golterman:2005ie,DelDebbio:2006cn,DelDebbio:2007pz,Farchioni:2004fs,Farchioni:2004us,Farchioni:2005tu} for lattice data).
\newline
In the infinite-volume limit also called thermodynamical limit flavor symmetry breaking can occur due to the fact that $\langle U_0 \rangle \neq 0$. Let us repeat here the analysis of \cite{Sharpe:1998xm} and apply it to our NLO results. We thus implicitly assume that the results obtained in the
$\epsilon$-regime pertain to this limit, as it was done e.g. in the analysis of \cite{Kieburg:2012fw}.
\newline
In order to determine the value of the minimum of the potential energy in eq. (\ref{ZNLOnf2})
we parameterize $U_0 = A + i B_j  \cdot \sigma_j$, with $\sigma_j$ the Pauli matrices. This makes the action in eq. (\ref{ZNLOnf2}) only depend on $A$.
If we assume that the sign of $c_2$ is positive the potential will be a parabola, and the minimum is given at LO by the parameter usually called
\begin{equation}\hat{\varepsilon} = \frac{m \Sigma}{16 a^2 c_2}
= \frac{\hm}{16 \ha^2 c_2}\ .
\end{equation}
\noindent
If this parameter lies outside the range $-1$ to $1$ then it is simple to see that the vector
symmetry can not be spontaneously broken, and the minimum is taken by $A=1$. However, if
the minimum satisfies $|\hat{\varepsilon}| < 1$ the vacuum is determined by $A^*$ = $\hat{\varepsilon}$. As a consequence
$B_{j}^* \neq 0$ and flavor symmetry is spontaneously broken to
$U(1)$. This tells us that the region $-1 < \hat{\varepsilon} < 1$ has the properties of the Aoki phase. We denote the value at which the transition takes place by $c_2^* = \frac{\hat{m}}{16 \hat{a}^2}$.\newline
One can repeat the same analysis using our NLO partition function to analyze the role of our corrections to this picture. The parameter $\hat{\varepsilon}$ is obviously modified at finite volume and lattice spacing and more precisely it is given by
\bea \hat{\varepsilon}_{V}
&=& \frac{\hm \Sigma^{\text{eff}}}{16 \ha^2 \Sigma c_{2}^{\text{eff}}}  \label{cc}\\
&\approx&
\frac{\hm}{16 \ha^2 c_2}\left(
1+\frac{5}{2F^2}\Delta(0)-\frac{\ha}{\hm\sqrt{V}}
\left(2\ha^2d_1+32\ha^2d_2-3\frac{d_2}{c_2}\right)
-\frac{\hm}{c_2\ha}\left(\frac{c_3}{\Sigma}+\frac{d_2}{4c_2}\right)
\frac{1}{\sqrt{V}}
\right)
\nn
\eea
\noindent
up to ${\cal O}\left(\frac{1}{V}\right)$.
Reintroducing the volume dependence from eq. (\ref{V-scale}), we
can derive the thermodynamical limit taking $V \rightarrow \infty$, with both $m \Sigma V$ and $a^2 V$ finite and much larger than one. In such limit the eq. (\ref{cc}) becomes
\be
\lim_{V\gg1}\hat{\varepsilon}_{V}=\frac{m \Sigma - 2\, a^3\, d_1 - 32\, a^3\, d_2}{16 a^2\, c_2 + 16 a m c_3
+ 4\frac{a m \Sigma d_2}{c_2}}\ ,
\ee
that can be matched after using some identities in our appendix A with the results found in \cite{Sharpe:2005rq} in the $p$-regime (for further discussions of the infinite volume limit see also \cite{Sharpe:2008ke}).
In general it is possible to compare the limit of the two regimes (the $\epsilon$-regime where $m_{\pi} L \gg 1$ and the $p$-regime where $m_{\pi} L \ll 1$ ) by approaching $m_{\pi} L \sim 1$ either from below or above in the respective scaling limit. Such comparison has been investigated in
\cite{PH,SH} where an agreement was found for the chiral condensate and
Dirac operator spectrum at NLO \cite{PH} and for the pseudo-scalar two-point
function at NLO \cite{SH} (see also the matching to LO between
\cite{Sharpe:1998xm} and \cite{Kieburg:2012fw}
for the minimal value of the pion mass). Our result adds a further
quantity to this list at NLO. Thus NLO corrections, that by assumption are small, even without changing the features of the system can shift its phase boundary by modifying the range of the minimum of the parabola determining the potential energy of the system.

\subsection{Partition Function for Generic Number of Flavors and Fixed Index}
\label{nu}

In this section we will consider Wilson Chiral Perturbation Theory with $N_f$ flavors. The situation becomes more complicated since additional terms are allowed in the
Lagrangian. Indeed the Wilson chiral Lagrangian for a generic number of degenerate quark flavors with mass $m$ reads at LO
\begin{eqnarray}
\nonumber \mathcal{L}_{\rm LO} &=& \frac{F^2}{4} \text{Tr}\left( \partial_{\mu} U \partial_{\mu} U^{\dagger} \right) - \frac{m \Sigma}{2} \text{Tr}\left( U + U^{\dagger} \right)  - \frac{z \Sigma}{2} \text{Tr}\left( U - U^{\dagger} \right)
\hspace{1.5cm}\\
 && + a^2 W_8 \text{Tr}\left( U^2 + U^{\dagger\, 2}\right)
+ a^2 W_6\, \text{Tr}\left( U + U^{\dagger\, }\right)^2 + a^2 W_7\, \text{Tr}\left( U - U^{\dagger\, }\right)^2,
\label{ttt0}\end{eqnarray}
\noindent
where we have introduced a source $z$ for the axial quark mass, and for the NLO \cite{Hansen:2011mc}\footnote{Note that compared to \cite{Hansen:2011mc}
 we have absorbed a factor of $2\Sigma/F^2$ into our $w_j$.} it reads
\begin{eqnarray}
\nonumber
\mathcal{L}_{\rm NLO} &=& \, a\, w_4\, \text{Tr}\left[ \partial_{\mu} U\, \,  \partial_{\mu} U^{\dagger\, }\right] \text{Tr}\, \left[  U + U^{\dagger\, }\right] +\, a\, w_5\, \text{Tr}\left[ \partial_{\mu} U\, \partial_{\mu} U^{\dagger\, }\, \left(  U + U^{\dagger\, }\right)\right]\\
\nonumber  &&- a\, m\, w_6 \,\left( \text{Tr}\left[ U + U^{\dagger\, }\right]\right)^2 - a\, m\, w_7 \, \left(\text{Tr}\left[ U - U^{\dagger\, }\right]\right)^2 - a\, m\, w_8 \, \text{Tr}\left[ U^2 + U^{\dagger\, 2}\right]\\
\nonumber &&+ a^3 x_1 \, \left(\text{Tr}\left[ U + U^{\dagger\, }\right]\right)^3 +\, a^3\, x_2  \left(\text{Tr}\left[ U - U^{\dagger\, }\right]\right)^2 \text{Tr}\left[ U + U^{\dagger\, }\right]
\nn\\
&&+\, a^3\, x_3\, \text{Tr}\left[ U^2 + U^{\dagger\, 2}\right] \text{Tr}\left[ U + U^{\dagger\, }\right]
+\, a^3\, x_4\, \text{Tr}\left[ U^2 - U^{\dagger\, 2}\right] \text{Tr}\left[ U - U^{\dagger\, }\right] \nn\\
&&+\, a^3\, x_5  \text{Tr}\left[ U^3 + U^{\dagger\, 3}\right]  +\, a^3\, x_6\, \text{Tr}\left[ U + U^{\dagger\, }\right].
\label{ttt}
\end{eqnarray}
\noindent
At this point one can proceed following the same steps used in the $N_f$=2 analysis, namely expand the action up
to $\mathcal{O}(\epsilon^2)$ and integrate out the pion fluctuations. Just for stylistic reasons we report our detailed computation in appendix \ref{B-nfnu}. Here we present only the final result and make some clarifications.
\newline
As we can see from both the LO and NLO Lagrangians (\ref{ttt0}) and (\ref{ttt}) a lot of terms appear compared to the usual $N_f=2$ case.
At LO there are three independent LECs, \emph{ i.e.} $W_6, W_7$ and $W_8$, and we recall again that in the simple case of two flavors they combine just in one coefficient called $c_2 = W_6 + W_8/2$ while $W_7$ doesn't enter in the computation. At NLO we have eleven new coefficients that are divided as follows. We have order $\mathcal{O}(p^2 a)$, $w_4$ and  $w_5$ written in the first line of eq. (\ref{ttt}). Again one can show that for the case of two flavors these two terms combine in one contribution which coefficient that we called $c_0= w_4 + w_5/2$. Then there are three terms of order $\mathcal{O}(m a)$ whose coefficients are $w_6$, $w_7$ and $w_8$, and the expressions are listed in the second line of the same equation. Again for two flavors $w_7$ doesn't contribute and the other terms form a combination that we called $c_3 = w_6 +w_8/2$ previously. And finally there are the more tedious contributions of order $\mathcal{O}(a^3)$ with six different terms and the relative six coefficient $x_i$ with $i$ running from one to six. In the case of two flavors, only two are independent and in particular the non-trivial combinations $d_1 = x_1 + x_3/2 + x_5/4$ and $d_2 = x_6 - 4 x_3 - 3 x_5 $ enter in the game, whereas $x_2$ and $x_4$ don't contribute.\newline
Since we will work at fixed index $\nu$
we define the projection to the following Fourier components:
\begin{equation}\mathcal{Z}^{\nu}
\equiv \int_{U(N_f)}[d_HU(x)]\ \det[U(x)^\nu]\ e^{-S}\ .
\end{equation}
\noindent
Taking the Fourier sum over all components will lead back to the $SU(N_f)$ integral. Note that after fixing the index the $N_f=2$ case has as many terms
in the Lagrangian as for general $N_f$, as there are no identities left to simplify it.
At LO this partition function can be evaluated as
\begin{eqnarray}
\nonumber  \mathcal{Z}_{\rm LO}^{\nu} &=&   {\cal N}
\int_{U(N_f)} d_H U_0\, \text{det}\left[U_0^{\nu}\right]\text{exp}\left[ \frac{m \Sigma V}{2} \text{Tr}\left[ U_0 + U_0^{\dagger} \right]
+\frac{z \Sigma V}{2} \text{Tr}\left[ U_0 - U_0^{\dagger}
\right]
\right. \\
 &&\left. - a^2V W_{8}\text{Tr}\left[ U_0^2 + U_0^{\dagger\, 2}\right]
 - a^2V W_{6 } \left(\text{Tr}\left[ U_0 + U_0^{\dagger\, }\right]\right)^2 - a^2 VW_{7}\, \left(\text{Tr}\left[ U_0 - U_0^{\dagger\, }\right]\right)^2 \right] ,\nn\\
\label{ZLOnu1}
\end{eqnarray}
\noindent
with expectation values defined by
\begin{equation}\langle\, F(U_0)\, \rangle_{\text{LO}}^\nu = \frac{1}{\mathcal{Z}^{\nu}_{\text{LO}}}\int_{U(N_f)} d_H U_0\, F(U_0)\,  \, \text{det}\, [U_0^{\nu}]\, \,  e^{- S^{(0)}_{U_0}} \ ,\label{ZLOnu}
\end{equation}
\noindent
using the corresponding action $ S^{(0)}_{U_0}$ from eq. (\ref{ZLOnu1}). The group integral eq. (\ref{ZLOnu}) is known explicitly and given in appendix \ref{C-W2pt}.
\newline
Since we have to take expectation values with respect to a $U(N_f)$ integral instead of $SU(N_f)$ we have derived new identities between the expectation values of the various terms in appendix \ref{A-unf}.
\newline
After expanding the action up to $\mathcal{O}(\epsilon^2)$ and integrating out the fluctuations it becomes non-trivial to reabsorb all the terms in such a way that the NLO Lagrangian can be expressed in terms of the LO one. Indeed we have to make use of three relations listed in the appendix \ref{A-unf} to find that
\begin{eqnarray}
\nonumber  \mathcal{Z}_{\rm NLO}^{\nu} &=&  \frac{{\cal N}"}{\cal N}\left(
\mathcal{Z}_{\rm LO}^{\nu}\left(\hm^{\text{eff}}, \hz^{\text{eff}},
\ha_{6}^{\text{eff}}, \ha_{7}^{\text{eff}}, \ha_{8}^{\text{eff}}\right)
+ X_1^{\text{eff}} \frac{2 \ha^3}{\sqrt{V}} \frac{\partial^2}{\partial\ha_6^2\partial\hm}
\mathcal{Z}_{\rm LO}^{\nu}\left(\hm,\hz, \ha_6, \ha_7, \ha_8\right)
\right.\\
&&+ X_2^{\text{eff}} \frac{2 \ha^3}{\sqrt{V}}
\frac{\partial^2}{\partial\ha_7^2\partial\hm}
\mathcal{Z}_{\rm LO}^{\nu}\left(\hm,\hz, \ha_6, \ha_7, \ha_8\right)
+ X_3^{\text{eff}} \frac{2 \ha^3}{\sqrt{V}}
\frac{\partial^2}{\partial\ha_6^2\partial\hm}
\mathcal{Z}_{\rm LO}^{\nu}\left(\hm,\hz, \ha_6, \ha_7, \ha_8\right)
\nn\\
&&\left.- 4X_5^{\text{eff}}
\frac{\partial^2}{\partial\hz\partial\hm}
\mathcal{Z}_{\rm LO}^{\nu}\left(\hm,\hz, \ha_6, \ha_7, \ha_8\right)\right) ,
\label{ZNLOnufinalmain}
\end{eqnarray}
where the explicit expressions for the renormalized
constants
$\hm^{\text{eff}}, \hz^{\text{eff}},\ha_{6}^{\text{eff}}$, $\ha_{7}^{\text{eff}}, \ha_{8}^{\text{eff}}
$, as well as for
$X_{1}^{\text{eff}}, X_{2}^{\text{eff}}, X_{3}^{\text{eff}},
X_{5}^{\text{eff}}$  are derived in the appendix \ref{B-nfnu}.
\newline
When we set the source $\hz=0$
the last effective coupling vanishes, $X_5^{\text{eff}}|_{z=0}=0$.
Also note if we were to set all the extra LECs to zero that contribute to the chiral Lagrangian at NLO (as it happens for SChPT in the next section), that is $x_{1,\ldots,6}=0=w_{6,7,8}$, then all $X_{1,2,3,5}^{\text{eff}}=0$ would equally vanish, and we could again write the NLO partition function as a LO one with the couplings renormalized through the one-loop corrections. Also the masses $m$ and $z$ would then be renormalized with the same effective LEC $\Sigma^{\text{eff}}$. This is the situation we find below in SChPT, and it is also true for the NLO finite-volume corrections in the continuum.

\subsection{Two-Point Correlation Functions for $N_f=2$}
\label{WChPT2pt}

In this subsection we will calculate the two-point correlators of the scalar and pseudoscalar current densities, in analogy to the continuum results in \cite{Hansen90,Damgaard:2001js,Damgaard:2002qe}.
They are defined respectively by
\bea
S_{0}(x)&=& \bar{\psi}(x) \psi(x)
\, \, \, \hspace{0.6cm} ,
S_{b}(x)= \bar{\psi}(x) t_b \psi(x)\ ,\\
P_{0}(x)&=& i \bar{\psi}(x)\gamma_5 \psi(x)\, \, \,\, ,
P_{b}(x)= i \bar{\psi}(x) t_b \gamma_5 \psi(x)\ ,
\eea
\noindent
at the first non-trivial order in the $\epsilon$-expansion. In the
expressions for the isovectors
the $t_b$ are proportional to the Pauli matrices $t_b = \frac{1}{2} \sigma_b$ for $b=1,2,3$.
Following the standard procedure, in order to calculate these quantities one has to introduce the Hermitian sources $s$ and $p$ in the partition function through the replacement $M \rightarrow M + s_0(x)+
s_b(x) t_b
+i p_0(x)+ i p_b(x) t_b$, and take the following functional derivatives:
\begin{equation}
\langle S_b(x) S_c(0)\rangle = \frac{1}{\mathcal{Z}}\frac{\delta^2}{\delta s_b(x) \delta s_c(0)} \mathcal{Z}[s,p]\mid_{s=p=0} \ ,
\end{equation}
\begin{equation}
\langle P_b(x) P_c(0)\rangle = \frac{1}{\mathcal{Z}}\frac{\delta^2}{\delta p_b(x) \delta p_c(0)} \mathcal{Z}[s,p]\mid_{s=p=0} \ .
\end{equation}
\noindent
Before starting we recall that the isovector scalar ($s_{b=1,2,3}(x)$)
and the isoscalar pseudoscalar ($p_0(x)$) densities, as a property of the $SU(2)$ theory, are vanishing at LO and also NLO and thus we will not consider these quantities.
\newline
The additional sources lead to the following modification of the
LO Lagrangian eq. (\ref{actionnf2}) ${\cal L}_{\rm LO}\to
{\cal L}_{\rm LO}+\delta {\cal L}_{\rm LO}$ with
\begin{equation}
\delta{\cal L}_{\rm LO} =
-s_0(x)\frac{\Sigma}{2} \text{Tr}\left[ U + U^{\dagger}\right]
+ip_b(x)\frac{\Sigma}{2} \text{Tr}\left[ t_b(U - U^{\dagger})\right]\ ,
\label{deltaLLO}
\end{equation}
as well as to the corresponding modification of the
NLO Lagrangian eq. (\ref{actionnf2NLO})
\begin{equation}
\delta{\cal L}_{\rm NLO} =
s_0(x)ac_3\left(\text{Tr}[ U + U^{\dagger}]\right)^2
-ip_b(x)ac_3\text{Tr}[ t_b(U - U^{\dagger})]\text{Tr}[ U + U^{\dagger}]\ .
\end{equation}
In addition to the continuum, the two-point functions are only known up to the renormalization constants $W_{S,P}$
that depend on the fields $S_b$ or $P_b$,
leading to the form
$(1+a W_S)\langle S_b(x) S_c(0)\rangle$, and likewise for the pseudoscalars (see \cite{Aoki:2009ri} for details about the renormalization
procedure).
\newline
Now we have to calculate the correlators at NLO in the $\epsilon$-expansion. First, we expand
the observables up to $\mathcal{O}(\epsilon^2)$ using the NLO action, in the second step we perform the Gaussian integration over the fluctuations.
Here we report only the results for $SU(2)$
while we show explicitly the full calculations for $U(N_f)$ at fixed index
in the appendix \ref{C-W2pt}.
\newline
The advantage of the fixed index averages is that we could use the compact integral representations for the LO partition function, eq. (\ref{detrel}) or
eq. (\ref{ZNf2nucompact}) for $N_f=2$, to obtain the NLO expressions from eq.
(\ref{ZNLOnufinalmain}). The logarithmic derivatives with respect to the corresponding couplings then generate all group averages given explicitly in appendix \ref{C-W2pt} in eqs.
(\ref{SSNfnu}), (\ref{PPNfnu}) from these LO and NLO partition functions.
\newline
For practical purposes however the disadvantage of fixed index at NLO is the large number of LECs to enter the expressions, that is for $N_f=2$ the 3 LECs $W_{6,7,8}$ from LO plus an additional 6 combinations  from NLO in eq. (\ref{ZNLOnufinalmain}). For this reason we have not attempted to plot the NLO two-point functions calculated at fixed index.
\newline
For the two-point function of the scalar current density we obtain
\begin{eqnarray}\langle S_0(x) S_0(0)\rangle &=&
\frac{(\Sigma^{\text{eff}})^2}{4} \left\langle \left(\text{Tr}[U_0+U_0^{\dagger\, }]\right)^2\right\rangle_{\rm NLO}
- \frac{\Sigma^2}{2 F^2} \left
\langle \text{Tr}\left[U_0^2 + U_0^{\dagger\, 2}\right]
- 4  \right\rangle_{\rm LO} \Delta(x)
\nn\\
&&
-  \frac{\ha \Sigma c_3}{\sqrt{V}} \left\langle \left(\text{Tr}[ U_0 + U_0^{\dagger}]\right)^3\right\rangle_{\rm LO} \ ,
\end{eqnarray}
\noindent
where the averages are now over constant matrices $U_0\in SU(2)$ with the NLO
or LO partition function eq. (\ref{ZNLOnf2}), respectively.
Apart from the averaging partition function the expression in the first line completely agrees with the continuum expression, see e.g. \cite{Akemann:2008vp}, after setting $N_f=2$ with $\text{Tr}\left[U_0 - U_0^{\dagger}\right]=0$ there.
\newline
For the pseudoscalar sector we have
\begin{eqnarray}\langle P_b(x) P_b(0)\rangle &=& - \frac{(\Sigma^{\text{eff}})^2}{8} \left\langle
\text{Tr}\left[U_0^2 + U_0^{\dagger\, 2}\right]
- 4  \right\rangle_{\rm NLO}
\nn \\
&&
+ \frac{\Sigma^2}{4 F^2} \left\langle
\frac{3}{4}\left(\text{Tr}\left[U_0 + U_0^{\dagger}\right]\right)^2 - \text{Tr}\left[U_0^2 + U_0^{\dagger\, 2}\right] + 4 \right\rangle_{\rm LO} \Delta(x)
\nn \\
&& + \frac{\ha c_3 \Sigma}{2\sqrt{V}} \left\langle
\left( \text{Tr}\left[U_0^2 + U_0^{\dagger\, 2}\right]
- 4 \right) \text{Tr}\left[U_0 + U_0^{\dagger\,}\right] \right\rangle_{\rm LO},
\label{PS2pt}
\end{eqnarray}
\noindent
where we have used the $su(n_f)$ completeness relation eq. (\ref{sunfcomplete}) after summing over $b=1,2,3$.
Once again the first two lines agree with the continuum expression for $N_f=2$, apart from the different average.
\newline
Let's start by plotting the zero-momentum correlator for  different values of the mass $m$ and at fixed values of $c_2$ and lattice
spacing $a$. More in detail we consider a hypercubic symmetric lattice with $N_L = N_T = 48$ ($\beta_1 = 0.140461$) and use $F$ = 90 MeV and $\Sigma$ = 250 MeV. In table 2. we list all parameters and numerical values of the LECs used in the plots
below for details on the explicit integrals used see  appendix \ref{BBB}). \newline
In figure 1. we plot the integrated two-point correlator, eqs. (\ref{PS2pt}) and (\ref{slope}), comparing our results with the continuum limit \cite{Hansen90,Damgaard:2001js,Damgaard:2002qe},
and with the results obtained by B\"ar, Necco and Schaefer \cite{Bar:2008th} (see also \cite{Shindler:2009ri}) that are valid in the GSM$^*$ regime, where finite lattice spacing effects only enter at NLO, compared to LO in our LCE counting.

\vspace{1cm}
\begin{center}
\begin{tabular}{|c|c|c|c|c|c|c|}
  \hline
  & $a$\tiny{(fm)} & $m$\tiny{(MeV)} & $c_2$\tiny{(GeV$^{4}$)} &
$d_1$\tiny{(GeV$^{4}$)} & $d_2$\tiny{(GeV$^{4}$)} &
$c_3$\tiny{(GeV$^{4}$)} \\ \hline
 Fig. 1. & 0.08 & 7, 2, 0.5 & 0.01 & $10^{-5}$ & $10^{-5}$ & $10^{-5}$       \\ \hline
 Fig. 2. & 0.12, 0.10 & 1 & 0.005 ... 0.050 & $10^{-5}$   & $10^{-5}$ & $10^{-5}$     \\  \hline
 Fig. 3. & 0.10 & 1 & -0.1 ...  0.1  & $10^{-2}$    & 0   &  0     \\  \hline
 \end{tabular}
\end{center}
\begin{center}\textbf{Table 2}. Numerical values of the lattice spacings, masses
and LECs used in the evaluation of the two-point function from appendix \ref{BBB}.\end{center}

\noindent
\begin{center}
\begin{figure}
\begin{center}
\epsfig{file=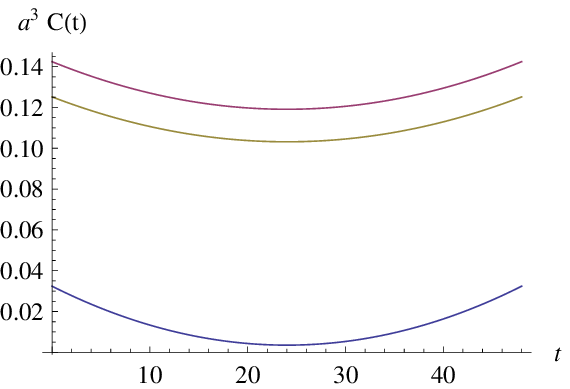,width=6.5cm}
\hspace{1cm}
\epsfig{file=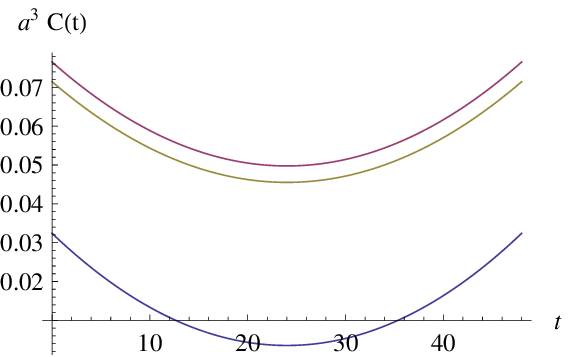,width=6.5cm}
\end{center}
\vspace{0.1cm}
\begin{center}\epsfig{file=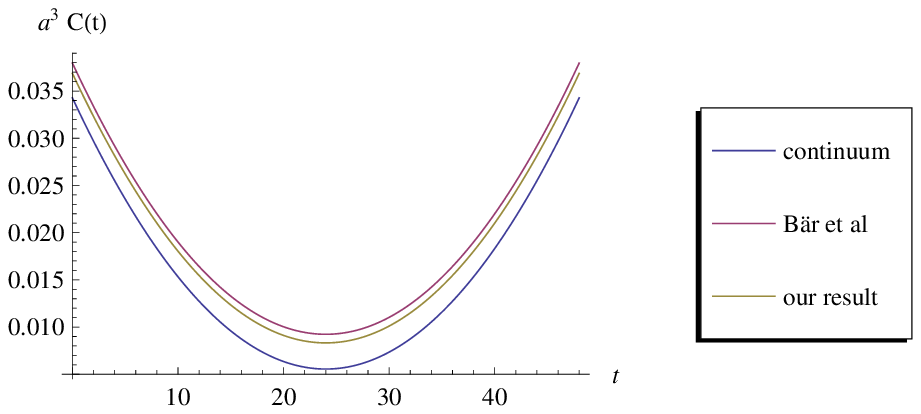,width=10cm}\end{center}
\caption{Pseudoscalar correlators at fixed $a$=0.08 and different masses $m$=0.5 (top-left), $m=2$ (top-right) and $m=7$ MeV (bottom). The different colors distinguish the continuum result, the results of \cite{Bar:2008th} in the GSM$^*$ regime and our result.}
\end{figure}
\end{center}
We only display the pseudoscalar-correlator, at three different masses with otherwise fixed parameters: for the first two values chosen, $m$=0.5 and $m$=2 MeV the system is in the Aoki regime (top two plots in fig. 1.) while for the last one with $m$=7 MeV it is in the GSM$^*$ regime instead  (bottom fig. 1).
Our calculation agrees quite well with \cite{Bar:2008th,Shindler:2009ri} when the GSM$^*$ counting is valid, while it disagrees when one enters the Aoki regime. Indeed the GSM$^*$ expansion is then no more reliable since lattice spacing effects give LO contributions and cannot be considered as perturbations. Similar plots could be obtained for the scalar two-point function.
\begin{center}
\begin{figure}
\epsfig{file=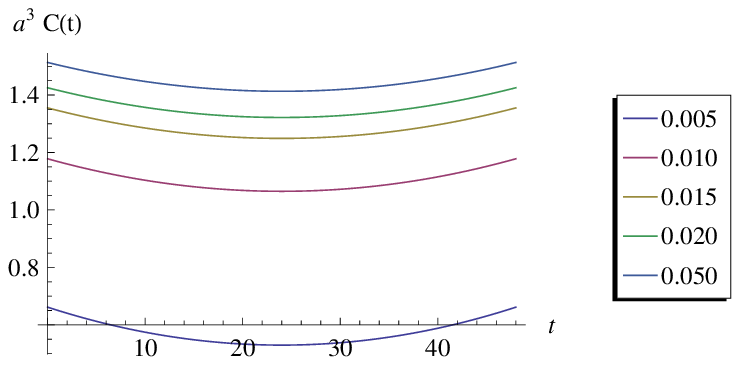,width=7.25cm}
\epsfig{file=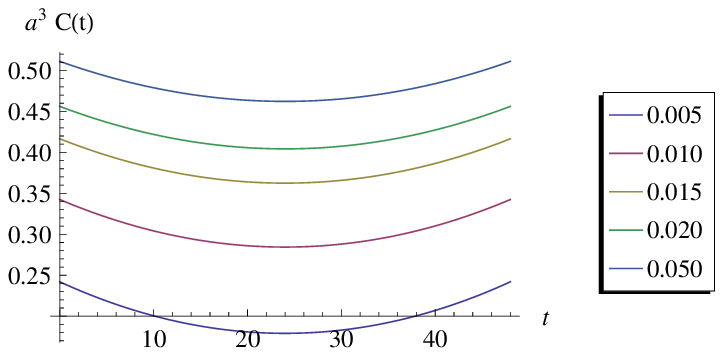,width=7.25cm}
\caption{Pseudoscalar correlators at fixed $a$=0.12 (left), $a=0.10$ (right) and fixed mass $m=1$ MeV, for different values of the LEC $c_2$ encoded by different colors.}
\end{figure}
\end{center}
In figure 2.  we plot the same pseudoscalar correlator for different values of  $c_2$ ranging from
0.005 to 0.05 GeV$^4$, at fixed $m$ and $a$. In the left figure we use a lattice spacing of $a=0.12$ fm while in the right plot we have $a$= 0.10 fm. As one could expect for the bigger lattice spacing small changes of $c_2$ bring up the parabola, making the corrections to the continuum more and more severe.
\newline
Finally we want to analyze the slope of the parabola described by the correlators. In order to do that we can recast them into the following form
given on the right hand side,
\begin{equation}
a^3 C(t)\equiv\int d^3x \sum_b \langle P^b(x,t) P^b(0)\rangle = A_P + B_P \left( \left( \left| \frac{t}{N_t} \right| - \frac{1}{2} \right)^2 - \frac{1}{24}\right)N_t a \ ,
\label{slope}
\end{equation}
\noindent
and study how the coefficient $B_P$ depends on the LECs. As we can see clearly from figure 2. the main correction to the continuum limit comes from the modification of the constant $A_P$ that determines the value of the minimum of the parabola. Indeed when the LEC $c_2$ increases (we assume that at NLO the other LECs have the same effect to increase $c_2^{\text{eff}}$), the minimum of the parabola becomes larger and larger compared to the continuum.
\newline

\begin{figure}[h]
\begin{center}
\epsfig{file=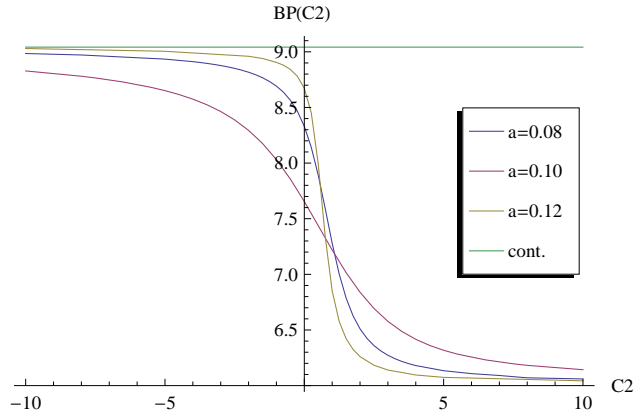,width=9cm}
\end{center}
\caption{The value of $B_P$ from eq. (\ref{slope}) as function of $c_2$. Both $x$- and $y$-axis are in units of $10^{10}$ MeV$^4$.}
\end{figure}
In figure 3. we look at the value of $B_P$, that is the coefficient that drives the slope of the parabola and thus is related to the masses of the pions.
Differently from what happens in the GSM$^*$ regime,
in the Aoki regime  also this parameter gets modified by lattice spacing effects. More
in detail, plotting the value of $B_P$ as a function of $c_2$ it is interesting to note that for small enough values of $c_2$ we are close to the continuum in the infinite-volume limit.
Referring back to our discussion of the Aoki phase boundary in subsection \ref{WChPTNf2},
for $-\infty < c_2 < c_2^*$
the system is in the so called Sharpe-Singleton scenario\footnote{The value $c_2^*$ is when $c_2$ satisfies $\hat{\epsilon}_V = 1$, see the discussion before eq. (\ref{cc}).}. In that region flavor symmetry is not broken and all three pions remain massive. The situation changes when $c_2$ is bigger than $c_2^*$, since the system enters in the Aoki phase. The value of $B_P$ decreases quite rapidly to reach a lower limit of
$2/3$ of the continuum limit. This is an indication that flavor symmetry is broken and as a consequence two of the three pions become massless, if we assume that every massless pion contributes equally to $1/3$ of the value of $B_P$.

\normalsize
\section{Staggered Chiral Perturbation Theory at NLO}\label{SChPT}
\subsection{Introduction}
\label{SChPTintro}
In this section we study finite-volume corrections to the LECs in the framework of SChPT. As we will see, the situation is easier compared to  WChPT since the corrections of order $\mathcal{O}( a\, m),\mathcal{O}(a^3)$ and $\mathcal{O}(a p^2)$  don't appear in the staggered Lagrangian. This simplifies remarkably the computation and allows us to write the expression for the NLO partition function with a generic number of flavors in terms of the LO one. Let us start to review briefly some basic known facts about SChPT and its equivalence to Staggered Chiral Random Matrix Theory (SChRMT) in the $\epsilon$-regime \cite{Osborn:2010eq}.
\begin{center}
\begin{tabular}{|c|c|}
  \hline
  Leading Order $\mathcal{O}(\epsilon^0)$ & $m$, $p^2$, $a^2$ \\
  Next-to-Leading Order $\mathcal{O}(\epsilon^2)$ &--\\
Next-to-Next-to-Leading Order   $\mathcal{O}(\epsilon^4)$ & $m^2$, $m p^2$, $p^4$, $a^2 m$, $a^2 p^2$, $a^4$ \\
  \hline
\end{tabular}
\end{center}
\begin{center}
\small
\textbf{Table 3.} Contributions to the Staggered Chiral Lagrangian in the Aoki regime.
\end{center}
\normalsize
\noindent
Staggered fermions are widely used to simulate quarks on the lattice. Indeed this formulation presents some clear advantages as the fact that the continuum chiral symmetry is not completely broken and that it is quite inexpensive to simulate numerically. However it doesn't solve completely the doubling problem. For every physical flavor there are four taste states that are degenerate in the continuum but split at finite lattice spacing because the taste symmetry is broken.
\newline
The effective chiral Lagrangian that describes the staggered formulation including finite lattice size corrections has been introduced in \cite{LeeSharpe} for the one flavor case and generalized to multiple flavors in \cite{AubinBernard}. The authors added all $\mathcal{O}(a^2)$ terms
to the continuum Lagrangian that are compatible with the staggered symmetries. As usual the breaking of the chiral symmetry from $G = SU(4N_f)_R \times SU(4N_f)_L$ to $SU(4N_f)_V$ is associated with the existence of light (pseudo) Goldstone boson fields that are collected into a $4 N_f \times 4 N_f$ unitary matrix $U$. For example in the case $N_f=3$ the matrix $U \in SU(12)$ can be parameterize as
\begin{center} $U = \left(\begin{array}{ccc}
  u & \pi^{+} & K^{+} \\
  \pi^{-} & d & K^0 \\
  K^{-} & \bar{K}^0 & s
\end{array}\right)$
 \end{center}
\noindent
where $=u,\, \pi^+,\, K^+\, ...\, $ are the $4 \times 4$ matrices that take into account the taste degrees of freedom and that can be written in the Dirac basis as
$U = \sum_b U_b T_b$, with denoting $T_b = \{ \xi_5,\, i \xi_{\mu}\xi_{5}, i \xi_{\mu}\xi_{\nu},\, \xi_{\mu},\, \xi_{I}  \}$.
In the LCE regime the LO Lagrangian, which is of order $\mathcal{O}(p^2, m, a^2)$, reads as
\cite{AubinBernard}
\begin{center}
\begin{eqnarray}
\nonumber  \mathcal{L}_{\rm LO} &=& \frac{F^2}{8} \text{Tr}\left( \partial_{\mu} U \partial_{\mu} U^{\dagger} \right) - \frac{\Sigma}{4} \text{Tr}\left( M^{\dagger} U + U^{\dagger} M \right) - a^2 C_1 \text{Tr}\left( U \gamma_5 U^{\dagger\,} \gamma_5\right)\\
\nonumber  &&- a^2 \frac{C_3}{2} \sum_{\mu}\, \left[ \text{Tr}\left( U \gamma_{\mu} U \gamma_{\mu}\right) + h.c. \right] - a^2 \frac{C_4}{2}\, \sum_{\mu}\left[  \text{Tr}\left( U \gamma_{\mu 5 } U \gamma_{\mu 5}\right) + h.c.\right]\\
\nonumber &&- a^2 \frac{C_{2V}}{4}\, \sum_{\mu} \left[  \text{Tr}\left( U \gamma_{\mu}\right) \text{Tr}\left( U \gamma_{\mu }\right) + h.c. \right]
- a^2 \frac{C_{2A}}{4}\, \sum_{\mu} \left[  \text{Tr}\left( U \gamma_{\mu 5}\right) \text{Tr}\left( U \gamma_{\mu 5 }\right) + h.c. \right] \\
\nonumber
&&- a^2 \frac{C_{5V}}{4}\, \sum_{\mu} \left[  \text{Tr}\left( U \gamma_{\mu}\right) \text{Tr}\left( U^{\dagger} \gamma_{\mu }\right) \right]-  a^2 \frac{C_{5A}}{4}\, \sum_{\mu} \left[  \text{Tr}\left( U \gamma_{\mu 5}\right) \text{Tr}\left( U^{\dagger} \gamma_{\mu 5 }\right) \right],\\
&&- a^2 C_6\, \sum_{\mu < \nu} \text{Tr}\left( U \gamma_{\mu \nu } U^{\dagger} \gamma_{\mu \nu }\right)\ ,
\label{no}
\end{eqnarray}
\end{center}
\noindent
where as usual $F$ and $\Sigma$ are the pion decay constant and the chiral condensate, respectively,  while the $4 N_f \times 4 N_f$ matrices $\gamma_{\mu}$  are the generalizations of the ordinary $4 \times 4$ Dirac matrices $\xi_{\mu}$ (see \cite{AubinBernard} for details)\footnote{There is also a mass term for the taste singlet pion that we have dropped.}. In addition to the continuum Gasser-Leutwyler Lagrangian there are some taste-breaking contributions and as a consequence some new LECs usually denoted as $C_1, C_{2\, A}, C_{2\, V}, C_3, C_4, C_{5\, A}, C_{5\, V}, C_6$.  For the one-flavor case the situation simplifies since all the two-trace terms in eq. (\ref{no}) can be Fierz transformed into one-trace terms (see \cite{LeeSharpe} for details). As we can see from table 3., if we want to go beyond LO, at NLO we will only get effects from one-loop finite-volume corrections of order $\mathcal{O}(\epsilon^2)$. Further LECs
arising from the discretization effects of order  $\mathcal{O}(a^4)$, $\mathcal{O}(a^2 p^2)$ and $\mathcal{O}(a^2 m)$ will only appear at NNLO, being of order $\mathcal{O}(\epsilon^4)$.
The corresponding terms are carefully listed in \cite{Sharpe:2004is}.
\newline
Quite recently it has been shown in \cite{Osborn:2010eq} that in the $\epsilon$-regime SChPT  is equivalent to SChRMT, including all one- and two-trace terms. In the latter theory the taste braking terms are introduced by adding a taste diagonal matrix to the usual Dirac operator
\begin{equation}\mathcal{D} = \left(
                                \begin{array}{cc}
                                  0 & i W \\
                                  i W^{\dagger} & 0 \\
                                \end{array}
                              \right) \otimes \mathbb{I}_4 + a^2 \mathcal{T}\ ,
\end{equation}
\noindent
with $W$ a random matrix of size $(N+\nu)\times N$ with complex entries, and where the explicit form of $\mathcal{T}$ and its relations with all the terms in the taste breaking potential are given in table 1. of \cite{Osborn:2010eq}. This correspondence is nothing else than the analogous relation for the staggered case between the $\epsilon$-regime of WChPT and WChRMT analyzed previously, valid at LO in the $\epsilon$-expansion.
The aim of this section is to study the finite-volume corrections to the LECs $C_i$ following the same
procedure utilized in the previous section \ref{WChPT} for WChPT.
As a byproduct of the calculation we will find how the taste splittings, \emph{i.e.} the difference between the mass squares of the non-Goldstone bosons and the Goldstone boson, are modified.

\subsection{Staggered Chiral Partition Function at $\mathcal{O}(\epsilon^2)$ for General $N_f$}
\label{SChPTnlo}

In order to calculate the partition function at $\mathcal{O}(\epsilon^2)$ in the staggered case we follow the same steps used
in the analysis of the Wilson chiral Lagrangian. Let us start to
rewrite the partition function as
\begin{equation}\mathcal{Z} = \int_{SU(4 N_f)} d_H U(x)\, e^{-S} = \int_{SU(4 N_f)} D_H U_0\, \, e^{-S_{U_0}}\, \, \mathcal{Z}_{\xi}(U_0)\ ,
\end{equation}
\noindent
where we have divided as usual the integration over the zero-modes $U_0$ from the integration over the fluctuations $\xi$ and where
\begin{equation}\mathcal{Z}_{\xi}(U_0) = \int d\xi(x)\, \left( 1 - \frac{2}{3 F^2 V} \int d^4x\, \text{Tr}\left[\xi(x)^2\right]\right) e^{S_{U_0} - S}.
\end{equation}
\noindent
Expanding the function $Z_{\xi}(U_0)$ up to order $\mathcal{O}(\epsilon^2)$ one can perform the Gaussian integrals over the fluctuations (for details about the expansion and the integration see appendix \ref{D-SChPT}). The next step is to reabsorb the finite-volume corrections into the LO Lagrangian by re-exponentiating all the terms found in the appendix \ref{D-SChPT}. At the end we can conclude that the NLO order partition function can be rewritten as the LO partition function with some renormalized LECs
\begin{equation}
\mathcal{Z}_{\rm NLO} =  \frac{{\cal N}'}{\cal N}\mathcal{Z}_{\rm LO}\left(\Sigma^{\text{eff}}, C_{i}^{ \text{eff}}\right).
\end{equation}
\noindent
These renormalized LECs following from the calculation in appendix \ref{D-SChPT}  are shown in table 4.
Similarly to the Wilson case the renormalized effective constants depend on the geometry of the system through the propagator $\Delta(0)$. As a further consequence we have extended the equivalence between SChPT and SChRMT up to NLO, as the form of the LO partition function is preserved for any number of flavors.

\vspace{0.2cm}
\begin{center}
\begin{tabular}{|c|c|}
  \hline
     &         \\
  \begin{math}\Sigma^{\text{eff}} =  \Sigma \left( 1 - \frac{16 N_f^2 - 1 }{4 F^2 N_f} \Delta(0)\right)\end{math} & $C_1^{\text{eff}}= C_1 \left( 1 - \frac{8 N_f}{F^2} \Delta(0)\right)$ \\
          &       \\
  $C_{2V}^{\text{eff}}=  C_{2V} - \frac{ C_{2V}(16 N_f^2 -2)+ 4 C_{3} N_f}{2 N_f\, F^2}\, \Delta(0)$  &
  $C_{2A}^{\text{eff}}= C_{2A} - \frac{C_{2A}(16 N_f^2 -2)+ 4 C_{4} N_f}{2 N_f\, F^2}\, \Delta(0)$ \\
                &        \\
  $C_3^{\text{eff}}= C_3  -  \frac{C_3(16 N_f^2 -2)+ 2\, C_{2V} N_f}{2 N_f\, F^2}\, \Delta(0) $ &
   $C_4^{\text{eff}}= C_4 -  \frac{C_4(16 N_f^2 -2)+ 2\, C_{2A} N_f}{2 N_f\, F^2}\, \Delta(0) $ \\
             &              \\
   $C_{5V}^{\text{eff}}= C_{5V} \left( 1 - \frac{8 N_f}{F^2} \Delta(0)\right)$ & $C_{5A}^{\text{eff}}= C_{5A} \left( 1 - \frac{8 N_f}{F^2} \Delta(0)\right)$ \\
           &        \\
   $C_6^{\text{eff}}= C_6 \left( 1 - \frac{8 N_f }{F^2} \Delta(0)\right)$ &   \\
               &             \\
   \hline
\end{tabular}
\end{center}
\begin{center}
\small
\textbf{Table 4.} The renormalized LECs in SChPT.
\end{center}
\vspace{0.2cm}
\normalsize
\noindent
 From the previous computation one can immediately understand how and if the finite-volume corrections affect the taste symmetry. Usually to study the taste symmetry violation one looks at the taste splitting $\Delta_{\xi_B}$ in the pion sector, \emph{i.e.} the difference between the mass square of a non-Goldstone pion and of the Goldstone one. At LO this quantity can be derived from a tree level expansion of the chiral Lagrangian, and indeed the masses of the non-neutral meson\footnote{For flavor neutral mesons the situation is more complicated and other terms have to be introduced in the chiral Lagrangian.}  composed of quark $b$ and $c$ can be written as
\noindent
\begin{equation}m^2 = \mu ( m_b + m_c ) + a^2 \Delta_{\xi_B}\ .
\end{equation}
\noindent
The terms $\Delta_{\xi_B}$ are related to the LECs through the relations
\cite{AubinBernard}
\begin{equation}\Delta_{P} = 0
\ ,\label{a1}\end{equation}

\begin{equation}\Delta_{A} = \frac{16}{F^2} \left( C_1 + 3 C_3 + C_4 + 3 C_6 \right)\ ,
\label{a2}\end{equation}

\begin{equation}\Delta_{T} = \frac{16}{F^2} \left( 2 C_3 + 2 C_4 + 4 C_6 \right)
\ ,\label{a3}\end{equation}

\begin{equation}\Delta_{V} = \frac{16}{F^2} \left( C_1 + C_3 + 3 C_4 + 3 C_6 \right)
\ ,\label{a4}\end{equation}

\begin{equation}\Delta_{I} = \frac{16}{F^2} \left( 4 C_3 + 4 C_4 \right)\ .
\label{a5}\end{equation}
\vspace{0.2cm}

\noindent
These splittings concern the Pseudoscalar ($P$), Axial-Vector ($A$), Tensor ($T$), Vector ($V$) and Singlet ($I$) taste pions respectively.
Since the LECs are modified at finite-volume the taste splitting get modified as follows
\begin{equation}\Delta_{P}^{\rm NLO} = \Delta_{P} = 0
\ ,\end{equation}

\begin{equation} \Delta_A^{\rm NLO} = \Delta_A -  \frac{16}{F^4} \left( 8 N_f [ C_1 + 3 C_6] + \frac{[C_4 + 3 C_3](16 N_f^2 -2)+ 2\, [ 3 C_{2V} + C_{2A} ] N_f}{2 N_f\,}\right)\Delta(0)
\ ,\end{equation}

\begin{equation}\Delta_T^{\rm NLO} = \Delta_T - \frac{16}{F^4} \left( 32 N_f C_6 +  \frac{[C_3 + C_4](16 N_f^2 -2)+ 2\, [ C_{2V} + C_{2A} ] N_f}{N_f\,} \right)\Delta(0)
\ ,\end{equation}

\begin{equation}\Delta_V^{\rm NLO} = \Delta_V -  \frac{16}{F^4} \left( 8 N_f [ C_1 + 3 C_6] + \frac{[3 C_4 + C_3](16 N_f^2 -2)+ 2\, [ C_{2V} + 3 C_{2A} ] N_f}{2 N_f\,}\right)\Delta(0)
\ ,\end{equation}

\begin{equation}\Delta_I^{\rm NLO} = \Delta_I - \frac{32}{F^4} \left(  \frac{[ C_3 + C_4 ](16 N_f^2 -2)+ 2\, [ C_{2V} + C_{2A} ] N_f}{ N_f\,} \right)\Delta(0)
\ .\end{equation}

\section{Summary and Discussion}
\label{summary}
In this paper we have computed ${\cal O}(\epsilon^2)$ finite-volume corrections in the so-called $\epsilon$-regime as they arise in Wilson and Staggered Chiral Perturbation theory. Thus we have taken into account both corrections to infinite-volume and to the continuum limit.
In SChPT ${\cal O}(a^2)$ effects only enter at LO, parameterized through a large number of in total 6 new low-energy constants. In contrast in WChPT such ${\cal O}(a^2)$ effects enter both at LO and at NLO in the $\epsilon$-expansion, leading to a total of 3 plus 9 LECs, respectively.
\newline
In consequence SChPT, although more complicated at LO, will remain simpler at NLO. In particular on the level of partition function the effect of NLO can be entirely expressed by renormalizing the LO LECs with corrections that we explicitly computed. As a second consequence the known equivalence between SChRMT and SChPT continues to hold at NLO. The drawback of the staggered formulation however remains, that the corresponding SChRMT has not been solved analytically to date. In addition our results for the NLO LECs provide us with the finite-volume corrections to the taste splittings as an application.
\newline
Turning to the Wilson case much is known about the LO spectrum of the Wilson Dirac operator at fixed index, due to the equivalent WChRMT picture. This equivalence breaks down at NLO for an arbitrary number of flavors  at fixed index including $N_f=2$, as extra derivatives appear when trying to express the NLO partition function through the LO partition function with renormalized, effective couplings. The reason is that we don't have enough group integral identities to absorb the high number of new terms at NLO into renormalised LO LECs.
Only for the original zero-mode group integral without fixing the index, and for the special case of $SU(2)$, the NLO partition function keeps its functional form compared to LO, and NLO gets absorbed into effective couplings. This is because here the number of group identities and of the new terms to be absorbed matches.
As a consequence we can use the two corresponding effective couplings to quantify
finite-lattice spacing effects on
the Aoki-phase transition in the thermodynamical limit.
For the same reason the finite-volume and ${\cal O}(a^2)$ corrections remain simplest for the two-point functions for $SU(2)$, which we have computed
in the scalar and pseudoscalar case
explicitly, and plotted in the pseudo-scalar sector for illustration.
\newline
Let us comment on finite-volume corrections to the positivity constraints on individual and certain combinations of LECs. At LO these were based on the positivity of the partition function at fixed index \cite{Akemann:2010em}, on Hermiticity arguments for the generating functional for the spectral density of Wilson Dirac eigenvalues \cite{Kieburg:2012fw}, and on the mass split using partially quenched WChPT  \cite{Hansen:2011kk}. It appears that neither line of argument can be easily translated to NLO, by simply replacing the LECs by effective ones. This has to do with the fact that at fixed index and/or for $N_f>2$ the functional form of the NLO partition function changes compared to LO.
\newline
In principle, the effect of NLO on the Wilson Dirac spectrum could be computed in the standard way, by introducing graded or replicated partition functions as generating functionals. However, due to the loss of determinantal structure of the partition function at NLO, that is observed at fixed index and certain vanishing couplings at LO (see appendix \ref{WChPT2pt} where we also computed an extended version), and due to the loss of the WChPT-WChRMT relation,
this seems to be a formidable task. Such a result would be very interesting in order to explain asymmetric effects on the spectrum attributed to NLO corrections in \cite{Damgaard:2011eg,Damgaard:2012gy}.
\newline
In principle other sources could be considered,
for example to compute vector and axial-vector two-point correlation functions.
Although we expect that the extension of our results is straightforward,
both for $SU(N_f)$ and for $U(N_f)$ at fixed index,
it is difficult to predict, if NLO effects can be absorbed by renormalising the LECs even for the special case $SU(2)$. This is due to the fact that the sources enter by shifting the kinetic term, rather than the mass term as for the scalar sources that we considered.
\newline
In any case the practical difficulty for the general results
we obtained for $SU(N_f>2)$ or for fixed index
is that a large number of effective LECs has to be determined
from actual data.\\

\noindent
{\bf Acknowledgments:}
We would like to thank Oliver B\"ar, Kim Splittorff and Edwin Laermann for fruitful discussions.
Partial support by the SFB$|$TR12 ``Symmetries and Universality
in Mesoscopic Systems'' of the German research council DFG is acknowledged (G.A.).
F.P. thanks the G. Galilei Institute for Theoretical
Physics in Florence for the hospitality.
F.P. is supported by the Research Executive Agency (REA)
of the European Union under Grant Agreement PITNGA-
2009-238353 (ITN STRONGnet).

\appendix
\section{Zero Mode Group Integral Identities}\label{Agroup-id}
\subsection{General $SU(N_f)$ Case} \label{A-sunf}

In order to derive some $SU(N_f)$ group identities among the expectation values of the various trace terms we follow the strategies
adopted in \cite{Damgaard:2001js,Akemann:2008vp}. We introduce the differentiation with respect to the group elements $U_{kl}$ of
$SU(N_f)$ defined as
\begin{equation}\nabla_b \equiv i ( t_{b} U )_{kl} \frac{\partial}{\partial U_{kl}}
\ ,\end{equation}
\noindent
where $t_b$ are the generators of the algebra $su(N_f)$
that satisfy the completeness relation
\begin{equation}
\label{sunfcomplete}
( t_b )_{ij} ( t_b )_{kl} = \frac12 \left(\delta_{il} \delta_{jk}
- \frac{1}{N_f}\delta_{ij} \delta_{kl}\right)
.\end{equation}
\noindent
This leads to the following derivatives
\begin{equation}
\nabla_b U=it_bU\ ,\ \ \nabla_b U^\dag=-iU^\dag t_b\ .
\label{delU}
\end{equation}
Considering that the Haar measure is left invariant, the integrals over total derivatives with respect to $\nabla_b$ have to vanish and thus for example
\begin{eqnarray}
\nonumber
0=\int_{SU(N_f)} d_H U\, \nabla_c \left\{\text{Tr}[ t_c G(U)]\, \text{exp}\left(\frac{m \Sigma V}{2} \text{Tr}\left[ U + U^{\dagger}\right]  -a^2 W_6 V \text{Tr}\left(\left[ U + U^{\dagger}\right]\right)^2  \right.\right. \\
 \left. \left. -\, a^2 W_7 V \text{Tr}\left(\left[ U - U^{\dagger}\right]\right)^2 - a^2 W_8 V\text{Tr}\left[ U^2 + U^{\dagger\, 2}\right]\right)\right\}\ \ \ \ \
\end{eqnarray}
\noindent
holds for any choice of the function $G(U)$. Throughout this appendix $U=U_0$ is a constant matrix and for simplicity we drop the subscript compared to the main text.
The following brackets denote the expectation value with respect to the integrand
\bea
&&\langle F(U)\rangle=1/{\cal Z}\\
&&\int_{SU(N_f)} d_H U\,
F(U)
e^{\frac{m \Sigma V}{2} \text{Tr}\left[ U + U^{\dagger}\right] - a^2 W_6 V \text{Tr}\left(\left[ U + U^{\dagger}\right]\right)^2 -  a^2 W_7 V \left(\text{Tr}\left[ U - U^{\dagger}\right]\right)^2 - a^2 W_8 V\text{Tr}\left[ U^2 + U^{\dagger\, 2}\right]}.\nn
\eea
Choosing $G(U)$ = $U - U^{\dagger}$ we obtain the following identity
\begin{eqnarray}
\nonumber
0&=&\left(N_f-\frac{1}{N_f}\right) \langle \text{Tr}\left[ U + U^{\dagger}\right]\rangle
+\frac{m \Sigma V}{2}\left(\langle \text{Tr}\left[ U^2 + U^{\dagger\, 2}\right]\rangle -2 N_f-\frac{1}{N_f}\langle \left(\text{Tr}\left[ U - U^{\dagger}\right]\right)^2\rangle\right)
\\
 \nonumber
&&- 2 a^2 W_6 V
 \left\langle \text{Tr}\left[ U + U^{\dagger}\right]
\left(\text{Tr}\left[ U^2 + U^{\dagger\, 2}\right]\rangle
-2N_f
-\frac{1}{N_f}
\left(\text{Tr}\left[ U - U^{\dagger}\right]\right)^2
\right)
\right\rangle
\\
&& - 2 a^2 W_7 V
\left(\langle \text{Tr}\left[ U^2 - U^{\dagger\, 2}\right]
\text{Tr}\left[ U - U^{\dagger}\right]\rangle
 -\frac{1}{N_f}\langle \left(\text{Tr}\left[ U - U^{\dagger}\right]\right)^2
 \text{Tr}\left[ U + U^{\dagger}\right]\rangle
 \right) \nn \\
&&  - 2 a^2 W_8 V \left(\langle \text{Tr}\left[ U^3 + U^{\dagger 3}\right]\rangle
  -\langle\text{Tr}\left[ U + U^{\dagger}\right]\rangle
  -\frac{1}{N_f}\langle \text{Tr}\left[ U^2 - U^{\dagger\, 2}\right]\text{Tr}\left[ U - U^{\dagger}\right]\rangle
  \right).\nn \\
  \label{Seconda}
\end{eqnarray}
\noindent

\subsection{The $SU(N_f = 2)$ Case}
\label{A-su2}
\normalsize
Here we rewrite the previous identity for the particular and more simple case of $N_f=2$ quarks. Indeed in this case some relations between the trace terms can be used to
simplify considerably
what we have found in (\ref{Seconda}). More in detail for any matrix $U$ that belongs to the group $SU(2)$ the following relations are valid
\begin{eqnarray}
&&\text{Tr}[ U - U^{\dagger}] = 0 \, \, \hspace{1cm}\, \, \ \text{Tr}[ U^2 + U^{\dagger\, 2}] = \frac{1}{2}\left(\text{Tr}[ U + U^{\dagger}]\right)^2 - 4\ ,\nn\\
&&\text{Tr}[ U^3 + U^{\dagger\, 3}] = \frac{1}{4}\left(\text{Tr}[ U + U^{\dagger}]\right)^3 -3
 \text{Tr}[ U + U^{\dagger}]\ .
\end{eqnarray}
\noindent
Note that these identities hold without taking an expectation value.
Thus one can see that the general $N_f$ expression of the identity
(\ref{Seconda})
reduces for this case to
\bea
0&=&\left(\frac{3}{2} + 16 a^2 c_2 V \right) \langle  \text{Tr}\left[ U + U^{\dagger}\right]\rangle + \frac{m \Sigma V}{4} \langle \left(\text{Tr}\left[ U + U^{\dagger }\right]\right)^2\rangle  - a^2 c_2 V \langle  \left(\text{Tr}\left[ U + U^{\dagger}\right]\right)^3\rangle \nn\\
&&- 4 m \Sigma V\ ,
\label{SecondaSU2}
\eea
that is the analogon of eq. (\ref{Seconda}), and we recall that we denoted by
$c_2=W_6+W_8/2$.

\subsection{General $U(N_f)$ Case} \label{A-unf}

In this part of the appendix we derive
$U(N_f)$ group identities that will be useful when working at fixed topology.
Differentiation with respect to the group elements $U_{kl}$ of $U\in U(N_f)$ is defined as before\,
\begin{equation}\nabla_b \equiv i ( t_{b} U )_{kl} \frac{\partial}{\partial U_{kl}}\ ,\end{equation}
\noindent
where $t_b$ are now the generators of the algebra $u(N_f)$ and satisfy the completeness relation
\begin{equation}
\label{unfcomplete}
( t_b )_{ij} ( t_b )_{kl} = \frac12 \delta_{il} \delta_{jk}.
\end{equation}
\noindent
Because of changing from $SU(N_f)$ to $U(N_f)$ group integrals we will have an extra factor $\det[U]^\nu$ included in the integrand, the derivative of which reads
\begin{equation}
\nabla_b \det[U]=i\Tr[t_b]\det[U]\ .
\label{deldetU}
\end{equation}
Again the integrals over total derivatives with respect to $\nabla_b$ have to vanish:
\begin{eqnarray}
\nonumber
0=\int_{U(N_f)} d_H U\, \nabla_c \left\{\text{Tr}[ t_c G(U)]\,
\det[U]^\nu \text{exp}\left(
\frac{m \Sigma V}{2} \text{Tr}\left[ U + U^{\dagger}\right]
+\frac{z \Sigma V}{2} \text{Tr}\left[ U - U^{\dagger}\right]
\right.\right. \\
 \left. \left.
- a^2 W_6 V \text{Tr}\left(\left[ U + U^{\dagger}\right]\right)^2
-\, a^2 W_7 V \text{Tr}\left(\left[ U - U^{\dagger}\right]\right)^2
- a^2 W_8 V\text{Tr}\left[ U^2 + U^{\dagger\, 2}\right]\right)\right\},\nn\\
\end{eqnarray}
\noindent
for any choice of $G(U)$. Here we have added an extra source term
$z$ for later
convenience. The brackets denoting expectation values
below
are now labeled
by the index $\nu$ in order to distinguish
them
from the previous subsection.
Also we use the following abbreviations:\newline
$m\Sigma V=\hm,\ z\Sigma V=\hz,\ \ha_j^2=a^2W_jV$ for $j=6,7,8$.
\newline
We now derive a series of identities. Consider
\noindent
\underline{$\bullet\ \ t_c=t_0\delta_{c,0}$ and $G[U]=1$:}
\begin{eqnarray}
0&=&\nu N_f
+\frac{\hm}{2} \langle \text{Tr}\left[ U - U^{\dagger}\right]\rangle^\nu
+\frac{\hz}{2} \langle \text{Tr}\left[ U + U^{\dagger}\right]\rangle^\nu
\nn\\
&&-2(\ha_6^2+\ha_7^2) \langle \text{Tr}\left[ U + U^{\dagger}\right]
\text{Tr}\left[ U - U^{\dagger}\right]\rangle^\nu
-2\ha_8^2  \langle \text{Tr}\left[ U^2 - U^{\dagger\,2}\right]\rangle^\nu\ ,
\label{Uid1}
\end{eqnarray}
\noindent
\underline{$\bullet\ \ t_c=t_0\delta_{c,0}$ and $G[U]=U-U^\dag$:}
\begin{eqnarray}
0&=&\langle \text{Tr}\left[ U + U^{\dagger}\right]\rangle^\nu
+\nu N_f\langle \text{Tr}\left[ U - U^{\dagger}\right]\rangle^\nu
+\frac{\hm}{2} \langle\left( \text{Tr}\left[ U - U^{\dagger}\right]\right)^2
\rangle^\nu
\nn\\
&&+\frac{\hz}{2} \langle \text{Tr}\left[ U + U^{\dagger}\right]
\text{Tr}\left[ U - U^{\dagger}\right]\rangle^\nu
-2(\ha_6^2+\ha_7^2) \langle \text{Tr}\left[ U + U^{\dagger}\right]
\left(\text{Tr}\left[ U - U^{\dagger}\right]\right)^2\rangle^\nu
\nn\\
&&-2\ha_8^2  \langle \text{Tr}\left[ U - U^{\dagger}\right]
\text{Tr}\left[ U^2 - U^{\dagger\,2}\right]\rangle^\nu\ ,
\label{Uid2}
\end{eqnarray}
\noindent
\underline{$\bullet$ summing over $t_c$ and $G[U]=U-U^\dag$:}
\begin{eqnarray}
0&=&N_f\langle \text{Tr}\left[ U + U^{\dagger}\right]\rangle^\nu
+\nu\langle \text{Tr}\left[ U - U^{\dagger}\right]\rangle^\nu
+\frac{\hm}{2} \langle\text{Tr}\left[ U^2 + U^{\dagger\,2}\right]
\rangle^\nu -\hm N_f
\nn\\
&&+\frac{\hz}{2} \langle
\text{Tr}\left[ U^2- U^{\dagger\,2}\right]\rangle^\nu
-2\ha_6^2\langle \text{Tr}\left[ U^2+ U^{\dagger\,2}\right]
\text{Tr}\left[ U + U^{\dagger}\right]\rangle^\nu
+4\ha_6^2N_f \langle\text{Tr}\left[ U + U^{\dagger}\right]\rangle^\nu
\nn\\
&&-2\ha_7^2\langle \text{Tr}\left[ U^2- U^{\dagger\,2}\right]
\text{Tr}\left[ U - U^{\dagger}\right]\rangle^\nu
-2\ha_8^2  \langle \text{Tr}\left[ U^3 + U^{\dagger\,3}\right]\rangle^\nu
+2\ha_8^2  \langle \text{Tr}\left[ U + U^{\dagger}\right]\rangle^\nu\ .\nn\\
\label{Uid3}
\end{eqnarray}
\noindent
We found two further identities with up to cubic powers of $U$ and $U^\dag$ by choosing  $G[U]=U+U^\dag$. However, these identities contain
new terms not present in the equations
we wish to simplify and hence they are not useful.

\section{Wilson Chiral Perturbation Theory for General $N_f$ at Fixed Index}
\label{B-nfnu}

In this section we focus on fixed index, the reason being that we then
can compute the group integrals more explicitly. Also we
have more group integral identities available from the previous subsection.
In this way we can express the NLO partition function through the LO
one at renormalized couplings, and derivatives thereof.
\newline
The partition function up to $\mathcal{O}(\epsilon^2)$ can be written as a sum of two contribution $S^{(0)}$
and $S^{(2)}$ that read
\small
\begin{description}
        \item[]
                  \begin{eqnarray} S^{(0)} &=& + \frac{1}{2}\mathrm{\int} d^4x \, \text{Tr}\left[\partial_{\mu} \xi \partial_{\mu} \xi\right] -  \frac{1}{2}\, m \Sigma V\, \text{Tr}\left[ U_0 + U_0^{\dagger} \right]
-\frac{1}{2}\, z \Sigma V\, \text{Tr}\left[ U_0 - U_0^{\dagger} \right]
\nn \\
            &&
+ a^2 V W_8 \text{Tr}\left[ U_0^2 + U_0^{\dagger\, 2}\right]
+\, a^2 V W_6\, \left(\text{Tr}\left[ U_0 + U_0^{\dagger\, }\right]\right)^2 + a^2 V W_7\, \left(\text{Tr}\left[ U_0 - U_0^{\dagger\, }\right]\right)^2\nn \\
               &\equiv& S_{\partial^2}^{(0)} +  S_{U_{0}}^{(0)}\ ,         \end{eqnarray}
         \item[]
        \begin{eqnarray}
        S^{(2)}& = & \frac{1}{12 F^2} \mathcal{\int} d^4x\, \text{Tr}\left[ [\partial_{\mu} \xi, \xi][\partial_{\mu} \xi, \xi]\right]  \nn\\
&&+     \frac{m\, \Sigma}{2 F^2}\, \mathcal{\int }d^4x \text{Tr}\left[ (U_0+ U_0^{\dagger})\xi^2  \right]
+     \frac{z\, \Sigma}{2 F^2}\, \mathcal{\int }d^4x \text{Tr}\left[ (U_0- U_0^{\dagger})\xi^2  \right]
\nn \\
        && -  2 a^2  \frac{W_8}{F^2} \mathcal{\int} d^4x\, \text{Tr}\left[\left( U_0^2 + U_0^{\dagger\, 2}\right) \xi^2\right] -  2 a^2  \frac{W_8}{F^2} \int d^4x\, \text{Tr}\left[ U_0 \xi U_0 \xi + U_0^{\dagger} \xi U_0^{\dagger} \xi\, \right]  \nn \\
       && - \, 2 a^2  \frac{ W_6 }{F^2} \int d^4x\, \left(\text{Tr}\left[ (U_0 - U_0^{\dagger\, }) \xi\right]\right)^2  - \, 2 a^2  \frac{ W_6 }{F^2} \int d^4x\, \text{Tr}\left[ (U_0 + U_0^{\dagger\, }) \xi^2\right]\text{Tr}\left[ U_0 + U_0^{\dagger}\right] \nn \\
       && - \, 2 a^2  \frac{ W_7 }{F^2} \int d^4x\, \left(\text{Tr}\left[ (U_0 + U_0^{\dagger\, }) \xi\right]\right)^2  - \, 2 a^2  \frac{ W_7 }{F^2} \int d^4x\, \text{Tr}\left[ (U_0 - U_0^{\dagger\, }) \xi^2\right]\text{Tr}\left[U_0-U_0^{\dagger}\right] \nn \\
       && + 2 a\frac{w_4}{F^2}\, \int d^4x\,  \text{Tr}\left[ \partial_{\mu} \xi \, \,  \partial_{\mu} \xi \right] \text{Tr}\, \left[  U_0 + U_0^{\dagger\, }\right] +\, 2 a \frac{ w_5}{F^2}\, \int d^4x\,   \text{Tr}\left[ \partial_{\mu} \xi\, \partial_{\mu} \xi\, \left(  U_0 + U_0^{\dagger\, }\right)\right]\nn \\
        && - a\, m\, w_6 V \,\left( \text{Tr}\left[ U_0 + U_0^{\dagger\, }\right]\right)^2 - a\, m\, w_7\, V\, \left(\text{Tr}\left[ U_0 - U_0^{\dagger\, }\right]\right)^2 - a\, m\, w_8 \, V\, \text{Tr}\left[ U_0^2 + U_0^{\dagger\, 2}\right] \nn \\
         && + a^3 x_1 \, V\, \left(\text{Tr}\left[ U_0 + U_0^{\dagger\, }\right]\right)^3 +\, a^3\, x_2\, V  \left(\text{Tr}\left[ U_0 - U_0^{\dagger\, }\right]\right)^2 \text{Tr}\left[ U_0 + U_0^{\dagger\, }\right] + \nn \\
         && + a^3\, x_3\, V\, \text{Tr}\left[ U_0^2 + U_0^{\dagger\, 2}\right] \text{Tr}\left[ U_0 + U_0^{\dagger\, }\right]  +\, a^3\, x_4\, V\, \text{Tr}\left[ U_0^2 - U_0^{\dagger\, 2}\right] \text{Tr}\left[ U_0 - U_0^{\dagger\, }\right]\nn \\
            && +\, a^3\, x_5\, V\,  \text{Tr}\left[ U_0^3 + U_0^{\dagger\, 3}\right]  +\, a^3\, x_6\, V\, \text{Tr}\left[ U_0 + U_0^{\dagger\, }\right].
        \end{eqnarray}
\end{description}
\noindent
\normalsize
They are the contributions of order $\mathcal{O}(\epsilon^0)$,
and $\mathcal{O}(\epsilon^2)$ respectively, where we have split the former one into the zero-mode and propagating mode part. The order $\mathcal{O}(\epsilon)$ vanishes.
\newline
Now we can proceed following the same steps used in the analysis of the two-flavor theory.
We begin
by rewriting the partition function for the $N_f$ flavors
with fixed index $\nu$ as
\begin{equation}
{\cal Z}^\nu = \int_{U(N_f)}\left[ d_H U\right]
\det[U_0]^\nu\, e^{-S} = \int_{U(N_f)} d_HU_0\, \det[U_0]^\nu
\, e^{-S^{(0)}_{U_0}}\, \, {\cal Z}_{\xi}(U_0)\ ,\end{equation}
\noindent
where
\begin{equation}
{\cal Z}_{\xi}(U_0) = \int_{SU(N_f)} \left[ d\xi(x) \right] \left( 1 - \frac{N_f}{3 F^2 V} \int d^4x\, \text{Tr}[\xi(x)^2]\right) e^{S_{U_0}^{(0)} - S}.\label{boss}\end{equation}
\noindent
The function ${\cal Z}_{\xi}(U_0)$ can be calculated
by
expanding  eq. (\ref{boss}) up to order $\mathcal{O}(\epsilon^2)$ and then performing the integral over the Gaussian fluctuation $\xi(x)$. One obtains
\small
\begin{eqnarray}
{\cal Z}_{\xi}(U_0) &=& \mathcal{N}\left\{ 1 -  \left( \frac{N_f^2-1}{N_f} \frac{m V \Sigma}{2 F^2} \Delta(0) + a^3 x_6 V\right) \text{Tr}\left[ U_0 + U_0^{\dagger} \right]
-\frac{N_f^2-1}{N_f} \frac{z V \Sigma}{2 F^2} \Delta(0)
\text{Tr}\left[ U_0 - U_0^{\dagger} \right]
\right. \nn \\
&& + \left(\frac{2 a^2 V}{F^2}\left( W_8 \frac{ N_f^2-2}{N_f} + W_6 + W_7\right) \Delta(0) + a m w_8 V \right) \text{Tr}\left[ U_0^2 + U_0^{\dagger\, 2}\right]
\nn \\
 && + \left( \frac{2 a^2 V}{F^2} \left( W_6 \frac{  N_f^2-1}{N_f} + \frac{N_f W_8- 2 W_7}{2 N_f}\right)\Delta(0) + a m w_6 V \right) \text{Tr}\left[ U_0 + U_0^{\dagger\, }\right]^2  \nn \\
 && + \left[ \frac{2 a^2 V}{F^2}\left( W_7 \frac{ N_f^2-1}{N_f} + \frac{N_f W_8- 2 W_6}{2 N_f}\right)  \Delta(0) + a m w_7 V \right] \text{Tr}\left[ U_0 - U_0^{\dagger\, }\right]^2 \nn \\
  && - a^3 x_1 \, V\, \left(\text{Tr}\left[ U_0 + U_0^{\dagger\, }\right]\right)^3 -\, a^3\, x_2\, V  \left(\text{Tr}\left[ U_0 - U_0^{\dagger\, }\right]\right)^2 \text{Tr}\left[ U_0 + U_0^{\dagger\, }\right] \nn \\
  &&  -\, a^3\, x_3\, V\, \text{Tr}\left[ U_0^2 + U_0^{\dagger\, 2}\right] \text{Tr}\left[ U_0 + U_0^{\dagger\, }\right]-\, a^3\, x_4\, V\, \text{Tr}\left[ U_0^2 - U_0^{\dagger\, 2}\right] \text{Tr}\left[ U_0 - U_0^{\dagger\, }\right] \nn \\
&& \left.-\, a^3\, x_5\, V\,  \text{Tr}\left[ U_0^3 + U_0^{\dagger\, 3}\right] \right\}.
\label{pooi} \end{eqnarray}
\normalsize
\noindent
Here $\mathcal{N}$ is a normalisation constant containing constants that are independent of $U_0$ and that can hence be pulled out of the integral.
Using the relations (\ref{Uid1})-(\ref{Uid3})  from the previous appendix \ref{A-unf} we can rearrange
several
terms present in eq. (\ref{pooi}) as a sum of other contributions. After several manipulation the final answer reads:

\scriptsize
\begin{eqnarray}
{\cal Z}_{\xi(U_0)} &=& \mathcal{N}`` \Big\{ 1  -\left[ \frac{N_f^2-1}{N_f} \frac{m V \Sigma}{2 F^2} \Delta(0)
+ \frac{a\,}{2 W_8} \left( x_5 N_f+x_4 - x_5 \frac{W_7}{W_8}
\right)
+ a^3V\left(x_5+  x_6 + \frac{2 x_5 N_f W_6}{ W_8} \right)
+ \frac{x_5 z^2 \Sigma^2V}{16 a W_8^2}
\right] \nn\\
&&
\times\text{Tr}\left[ U_0 + U_0^{\dagger} \right] 
\nn \\
&& - \left[ \frac{N_f^2-1}{N_f} \frac{z V \Sigma}{2 F^2} \Delta(0)+
\frac{a\nu}{2 W_8}\left(x_5
+ N_f\left( x_4\, - x_5 \frac{W_7}{W_8}\right) \right)
+ \frac{x_5 mz \Sigma^2 V}{16 a W_8^2}
\right] \text{Tr}\left[ U_0 - U_0^{\dagger} \right] \nn \\
&& + \left[\frac{2 a^2 V}{F^2}\left( W_8 \frac{ N_f^2-2}{N_f} + W_6 + W_7\right) \Delta(0) + a m w_8 V - \frac{a x_5 m \Sigma V }{4 W_8}\right] \text{Tr}\left[ U_0^2 + U_0^{\dagger\, 2}\right] \nn \\
&& \left. + \left[ \frac{2 a^2 V}{F^2} \left( W_6 \frac{  N_f^2-1}{N_f} + \frac{N_f W_8- 2 W_7}{2 N_f}\right)\Delta(0) + a m w_6 V \right] \left(\text{Tr}\left[ U_0 + U_0^{\dagger\, }\right]\right)^2 \right. \nn \\
&&   + \left[ \frac{2 a^2 V}{F^2}\left( W_7 \frac{ N_f^2-1}{N_f} + \frac{N_f W_8- 2 W_6}{2 N_f}\right)  \Delta(0)
+ a mV\left( w_7  - \left( x_4\, - x_5 \frac{W_7}{W_8}\right)\frac{\Sigma }{4 W_8} \right)
\right] \left(\text{Tr}\left[ U_0 - U_0^{\dagger\, }\right]\right)^2   \nn \\
&& -\, a^3\, V \left( x_2\, - \frac{W_6 + W_7}{W_8}\left( x_4\, - x_5 \frac{W_7}{W_8}  \right)\right) \left(\text{Tr}\left[ U_0 - U_0^{\dagger\, }\right]\right)^2 \text{Tr}\left( U_0 + U_0^{\dagger\, }\right)\nn \\
&& \left. -\, \, a^3\, V\,\left( x_3 - \frac{x_5 W_6}{W_8}\right) \text{Tr}\left( U_0^2 + U_0^{\dagger\, 2}\right) \text{Tr}\left( U_0 + U_0^{\dagger\, }\right) - a^3 x_1 \, V\, \text{Tr}\left( U_0 + U_0^{\dagger\, }\right)^3 \right.\nn\\
&& \left. -\left( x_4\, - x_5 \frac{(W_6+2W_7)}{W_8}\right) \frac{a z\Sigma V}{4 W_8}
\text{Tr}\left( U_0 - U_0^{\dagger\, }\right)  \text{Tr}\left( U_0 + U_0^{\dagger\, }\right) \right\}.
\label{pooi4}
\end{eqnarray}
\normalsize
\noindent
The modification of the normalisation constant denoted by $\mathcal{N}``$ results from the constant $U_0$-independent parts in the identities
(\ref{Uid1}) and (\ref{Uid3}) that we have employed.
At this point it turns out to be useful to define the new renormalized
masses  and LECs as
\bea
\hm^{\text{eff}}
&=& \hm-
\left[ \frac{(N_f^2-1)}{N_f} \frac{\hm}{2 F^2} \Delta(0)
+ \frac{\ha}{2 W_8\sqrt{V}} \left( x_4+x_5 \left( N_f- \frac{W_7}{W_8}\right)
\right)
\right.\nn\\
&&\left.
+ \frac{\ha^3}{\sqrt{V}}\left( x_6 +x_5\left(1+\frac{2 N_f W_6}{ W_8} \right) \right)
+ \frac{x_5 \hz^2}{16 \ha W_8^2\sqrt{V}}\right],
\eea
\bea
\hz^{\text{eff}}
&=& \hz-
\left[ \frac{(N_f^2-1)}{N_f} \frac{\hz}{2 F^2} \Delta(0)
+
\frac{\ha\nu}{2 W_8\sqrt{V}}\left( x_5+ N_f\left( x_4\, - x_5 \frac{W_7}{W_8}\right)
\right)
+ \frac{x_5 \hm\hz}{16 \ha W_8^2\sqrt{V}}\right]\!\!,\ \ \ \ \ \ \ \
\eea

\begin{equation}
(\ha_{8}^{\text{eff}})^2
= \ha_8^2  - \left[\frac{2 \ha^2}{F^2}\left( W_8 \frac{ N_f^2-2}{N_f}
+ W_6 + W_7\right) \Delta(0)
+ \frac{\ha \hm}{\sqrt{V}}\left( \frac{w_8}{\Sigma}  - \frac{x_5 }{4 W_8}\right)\right],
\end{equation}

\begin{equation}
(\ha_{6}^{\text{eff}})^2
= \ha_6^2 -\left[ \frac{2 \ha^2}{F^2} \left( W_6 \frac{  (N_f^2-1)}{N_f} + \frac{N_f W_8- 2 W_7}{2 N_f}\right)\Delta(0)
+ \frac{\ha \hm w_6}{\sqrt{ V}} \right],
\end{equation}

\bea
(\ha_{7}^{\text{eff}})^2
&=&  \ha_7^2 - \left[ \frac{2 \ha^2}{F^2}\left( W_7 \frac{ (N_f^2-1)}{N_f} + \frac{N_f W_8- 2 W_6}{2 N_f}\right)  \Delta(0)
\right.\nn\\
&&\left.
\ \ \ \ \ \
+ \frac{\ha \hm}{\sqrt{V}}
\left(
\frac{w_7}{\Sigma}  - \left( x_4\, - x_5 \frac{W_7}{W_8}\right)
\frac{1}{4 W_8 } \right)\right],
\eea

\begin{equation}X_{1}^{\text{eff}} =   x_1\ ,
\end{equation}

\begin{equation}X_{2}^{\text{eff}} =  \left( x_2\, - \frac{W_6 + W_7}{W_8}\left( x_4\, - x_5 \frac{W_7}{W_8}  \right)\right),   \end{equation}

\begin{equation}X_{3}^{\text{eff}} =   \left( x_3 - \frac{x_5 W_6}{W_8}\right),\end{equation}

\begin{equation}X_{5}^{\text{eff}} =
\frac{\ha \hz}{4 W_8\sqrt{V}}
\left( x_4\, - x_5 \frac{(W_6+2W_7)}{W_8}\right).
\end{equation}
\normalsize
Note that at NLO the quark and axial quark mass do not renormalize
with the same LEC $\Sigma$ any more. The renormalized constants
$\hm^{\text{eff}}, \hz^{\text{eff}},\ha_{6,7,8}^{\text{eff}}$ contain both
${\cal O}(1)$ and ${\cal O}(\epsilon^2)$ parts,
the constants $X_{1,2,3}^{\text{eff}}$ are all ${\cal O}(1)$ only, and
$X_{5}^{\text{eff}}$ is of  ${\cal O}(\epsilon^2)$ only.
This is because of the following NLO expression for the partition function:
\begin{eqnarray}
\nonumber  \mathcal{Z}_{\rm NLO}^{\nu} &=& \frac{{\cal N}"}{\cal N}\left( \mathcal{Z}_{\rm LO}^{\nu}\left(\hm^{\text{eff}}, \hz^{\text{eff}},
\ha_{6}^{\text{eff}}, \ha_{7}^{\text{eff}}, \ha_{8}^{\text{eff}}\right)
+ X_1^{\text{eff}} \frac{2 \ha^3}{\sqrt{V}} \frac{\partial^2}{\partial\ha_6^2\partial\hm}
\mathcal{Z}_{\rm LO}^{\nu}\left(\hm,\hz, \ha_6, \ha_7, \ha_8\right)
\right.\\
&&+ X_2^{\text{eff}} \frac{2 \ha^3}{\sqrt{V}}
\frac{\partial^2}{\partial\ha_7^2\partial\hm}
\mathcal{Z}_{\rm LO}^{\nu}\left(\hm,\hz, \ha_6, \ha_7, \ha_8\right)
+ X_3^{\text{eff}} \frac{2 \ha^3}{\sqrt{V}}
\frac{\partial^2}{\partial\ha_8^2\partial\hm}
\mathcal{Z}_{\rm LO}^{\nu}\left(\hm,\hz, \ha_6, \ha_7, \ha_8\right)
\nn\\
&&\left.- 4X_5^{\text{eff}}
\frac{\partial^2}{\partial\hz\partial\hm}
\mathcal{Z}_{\rm LO}^{\nu}\left(\hm,\hz, \ha_6, \ha_7, \ha_8\right)\right) .
\label{ZNLOnufinal}
\end{eqnarray}


\section{Scalar and Pseudoscalar Currents in WChPT at Fixed Index}
\label{C-W2pt}

In this appendix we will complement the main body of this paper by computing the partition function and scalar and pseudoscalar two-point functions for an arbitrary number of flavors $N_f$ at fixed index $\nu$. We begin with the
partition function. Given the previous appendix we only need to compute it to LO as the NLO one can be expressed through it. It is defined as
\bea
{\cal Z}_{\rm LO}^{N_f,\,\nu}(\hm,\hz,\ha_6,\ha_7,\ha_8)
&\equiv&\int_{U(N_f)}d_H U\ \det[U]^\nu \exp\left[
\frac{\hm}{2}\text{Tr}\left[ U + U^{\dagger} \right]
+\frac{\hz}{2}\text{Tr}\left[ U - U^{\dagger} \right]
\right.
\nn \\
            &&\left.
- \ha^2_8 \text{Tr}\left[ U^2 + U^{\dagger\, 2}\right]
- \ha^2_6\left(\text{Tr}\left[ U + U^{\dagger\, }\right]\right)^2
-\ha^2_7\left(\text{Tr}\left[ U - U^{\dagger\, }\right]\right)^2
\right],\nn\\
\eea
with
the number of quark flavors $N_f$ explicitly displayed,
and with the rescaled quantities $\hm=mV\Sigma$, $\hz=zV\Sigma$,
$a^2VW_j=\ha^2_j$ for $j=6,7,8$. We have dropped the index of $U_0$ here and in the following.
This integral has been calculated in the literature in a series of works \cite{Damgaard:2010cz,Akemann:2010em}. Let us briefly review and
slightly extend their results.
\newline
Consider the following group integral that contains the above case for
$\ha_6=\ha_7=0$:
\bea
{\cal I}^{N_f,\,\nu}
&\equiv&\int_{U(N_f)}d_H U\ \det[U]^\nu\exp\left[\sum_{j=1}^\infty
\left(\alpha_j\text{Tr}[U^j] +\beta_j \text{Tr}[U^{\dag\,j}]\right)\right].
\label{INfnu}
\eea
After diagonalizing the matrix, $U\to v\,$diag$(e^{i\theta_1},\ldots,e^{i\theta_{N_f}})v^\dag$, with $v\in U(N_f)/U(1)^{N_f}$, it can be written as a determinant over a single integral:
\bea
{\cal I}^{N_f,\,\nu}&=& {\cal C}_{N_f}\int_{-\pi}^{\pi}\prod_{l=1}^{N_f}
d\theta_l \exp[i\nu\theta_l]\exp\left[
\sum_{j=1}^\infty
\left(\alpha_je^{ij\theta_l} +\beta_j e^{-ij\theta_l}\right)
\right]\
\prod_{k>n}^{N_f}\left| e^{i\theta_k}-e^{i\theta_n}\right|^2
\nn\\
&=& {\cal C}_{N_f}\int_{-\pi}^{\pi}\prod_{l=1}^{N_f}
d\theta_l\ \
\det_{1\leq n,k\leq N_f}\left[ \exp\left[i(\nu+n-1)\theta_k+\sum_{j=1}^\infty
\alpha_je^{ij\theta_k}
\right]\right]\nn\\
&&\times\det_{1\leq n,k\leq N_f}\left[ \exp\left[-i(n-1)\theta_k+\sum_{j=1}^\infty
\beta_je^{ij\theta_k}
\right]\right]\nn\\
&=&{\cal C}_{N_f}N_f!\ \det_{1\leq n,k\leq N_f}\left[
\int_{-\pi}^{\pi}d\theta\ e^{i\theta(\nu+k-n)}
e^{\sum_{j=1}^\infty (\alpha_j\exp[ij\theta] +\beta_j \exp[-ij\theta])}
\right].
\eea
In the first step have rewritten the absolute value square of the Vandermonde determinant, the Jacobian resulting from the diagonalization, and pulled the exponential prefactors into the respective determinants. In the second step we have applied one of the de Bruijn integration formulas.
The constant ${\cal C}_{N_f}$ is the volume of the coset integral over $v$.
\newline
As a consequence for $\ha_6=\ha_7=0$ we can write the corresponding $N_f$-flavor partition function as a determinant of a single flavor partition function,
\begin{equation}
\mathcal{Z}_{\rm LO}^{N_f,\,\nu}(\hat{m}, \hat{z},0,0, \hat{a}_8)
\sim \text{det}\left[ \mathcal{Z}_{\rm LO}^{N_f=1,\,\nu+k-n}(\hat{m}, \hat{z},0,0, \hat{a}_8)\right]_{k,n=1,\ldots,N_f}\ ,
\label{detrel}
\end{equation}
\noindent
where
\begin{equation}
\mathcal{Z}_{\rm LO}^{N_f=1,\,\nu}(\hat{m}, \hat{z},0,0, \hat{a}_8)
= {\cal C}_1\int_{- \pi}^{\pi} \exp[i \theta \nu+ \hat{m} \cos{(\theta)} + i \hat{z} \sin{(\theta)} -2 \hat{a}_8^2 \cos{( 2 \theta)}]\ .
\end{equation}
\noindent
In the particular case of two degenerate flavors eq. (\ref{detrel}) reduces to
\begin{equation}\mathcal{Z}^{N_f=2,\,\nu} \sim \left(\mathcal{Z}^{N_f=1,\,\nu}\right)^2 - \mathcal{Z}^{N_f=1,\,\nu+1}\mathcal{Z}^{N_f=1,\,\nu-1}, \label{Nf2det}\end{equation}
\noindent
where we have suppressed the arguments. The remaining two LECs $\ha_{6,7}$ can be switched on by performing two Gaussian integrals on
the above formulas, following \cite{Akemann:2010em}.
In addition it has been shown in \cite{Kieburg:2012fw} based on Hermiticity that
both LECs have to be non-positive, $\ha_{6},\ha_7\leq 0$. Consequently we
obtain the following expression:
\begin{equation}\mathcal{Z}_{\rm LO}^{N_f,\,\nu}(\hat{m},\hat{z},\hat{a}_6,\hat{a}_7,\hat{a}_8)
= \int_{-\infty}^{\infty}
\frac{dy_6 dy_7 }{16 \pi |\hat{a}_6 \hat{a}_7|} \
e^{-\frac{y_6^2}{16 \hat{a}_6^2}} e^{-\frac{y_7^2}{16 \hat{a}_7^2}} \mathcal{Z}_{\rm LO}^{N_f,\,\nu}(\hat{m} - y_6,\hat{z} -y_7,0,0,\hat{a}_8)\, .
\end{equation}
\noindent
Now let's derive the partition function at fixed topology in the case of two flavors.  In fact the Gaussian integration over the partition function
(\ref{Nf2det}) can be performed explicitly, and we obtain that
\begin{eqnarray}\mathcal{Z}_{\rm LO}^{2,\nu}(\hat{m},\hat{z},\hat{a}_6,\hat{a}_7,\hat{a}_8)
&=&
\frac{{\cal C}_2}{2
}
\int_{-\pi}^{\pi} d\theta_1 d\theta_2\, e^{i (\theta_1 + \theta_2 ) \nu}\left( 1 - e^{i (\theta_1 - \theta_2 )}\right) e^{\hat{m} ( \cos{\theta_1} + \cos{\theta_2} )+i \hat{z} ( \sin{\theta_1} + \sin{\theta_2} ) }\nn \\
 &&\times
 e^{4 \hat{a}_6^{2} ( \cos{\theta_1} + \cos{\theta_2} )^2}
e^{-4 \hat{a}_7^{2} ( \sin{\theta_1} + \sin{\theta_2} )^2 }
e^{- 2 \hat{a}_8^{2} ( \cos{2 \theta_1} + \cos{2 \theta_2} ) }.
\label{ZNf2nucompact}
\end{eqnarray}
\noindent
Although in this expression the integrals do not factorize it is very useful for a numerical integration. The normalization constant ${\cal C}_{2}$ is not important as it drops out in expectation values.
\newline
We can now compute the NLO two-point functions for general $N_f$ at fixed index, following the same lines as in the previous appendix \ref{B-nfnu}.
In fact written in terms of group averages $\langle\ldots \rangle^\nu$, where the superscript denotes the index, the expressions with or without fixing the index don't differ. This is because the propagating modes that we contract always live in $SU(N_f)$. The only difference is that for fixed index at $N_f=2$ we no longer have the $SU(2)$ identities at hand, e.g.
$\text{Tr}[U-U^\dag]\neq0$ no longer applies.
\newline
For simplicity we will only present that unflavored scalar and flavored pseudoscalar two-point functions as in the main text. The results we obtain are
\bea
\langle S_0(x) S_0(0)\rangle^\nu &=&
\frac{(\Sigma^{\text{eff}})^2}{4} \left\langle \left(\text{Tr}[U+U^{\dagger\, }]\right)^2\right\rangle_{\rm NLO}^\nu
-  \frac{\ha \Sigma c_3}{\sqrt{V}} \left\langle \text{Tr}[ U + U^{\dagger}]^3\right\rangle_{\rm LO}^\nu \nn\\
&-& \frac{\Sigma^2}{2 F^2} \left\{
\left\langle \text{Tr}\left[U^2 + U^{\dagger\, 2}\right]\right\rangle_{\rm LO}^\nu
- 2N_f
- \frac{1}{N_f}\left\langle\left( \text{Tr}\left[U - U^{\dagger}\right]\right)^2 \right\rangle_{\rm LO}^\nu
\right\}\Delta(x) .
\nn\\
\label{SSNfnu}
\eea
For the flavored pseudoscalars we sum over all generators $t_b$ of
$su(N_f)$
\bea
\sum_b \langle P_b(x) P_b(0)\rangle^\nu &=&
-\frac{(\Sigma^{\text{eff}})^2}{8} \left\{
\left\langle
\text{Tr}\left[U^2 + U^{\dagger\, 2}\right]
\right\rangle_{\rm NLO}^\nu
-2N_f
- \frac{1}{N_f}\left\langle
\left( \text{Tr}\left[U - U^{\dagger}\right]\right)^2
\right\rangle_{\rm NLO}^\nu
\right\}\nn\\
&+& \frac{\ha c_3 \Sigma}{2\sqrt{V}} \left\langle
\left( \text{Tr}\left[U^2 + U^{\dagger\, 2}\right]
- \frac{1}{N_f}\left( \text{Tr}\left[U - U^{\dagger}\right]\right)^2
- 2N_f \right) \text{Tr}\left[U + U^{\dagger\,}\right]
\right\rangle_{\rm LO}^\nu
\nn \\
&+&
\frac{\Sigma^2}{4 F^2} \left\{
\frac{1}{2}\left\langle
\left(\text{Tr}\left[U - U^{\dagger}\right]\right)^2
\right\rangle_{\rm LO}^\nu
+
\frac{(N_f^2+2)}{2N_f^2}
\left\langle
\left(\text{Tr}\left[U + U^{\dagger}\right]\right)^2
\right\rangle_{\rm LO}^\nu\right.\nn\\
&&\ \ \ \ \ \ \ \ \ \ \left.+2N_f^2
- \frac{2}{N_f}
\left\langle
\text{Tr}\left[U^2 + U^{\dagger\, 2}\right]\right\rangle_{\rm LO}^\nu
\right\}\Delta(x)\ .
\label{PPNfnu}
\eea
\noindent
Note that
in deriving the expression for the zero-momentum correlation functions, which are functions only of the Euclidean time $t$, we would have to make use of the relation
\begin{equation}\int d^3 x\, \Delta(x-y) = \frac{a N_T}{2}\left[\left( \left|\frac{t_0}{T}\right| - \frac{1}{2}\right)^2  - \frac{1}{24} \right]. \end{equation}


\section{Explicit Computation of Partition Function and Currents for $SU(2)$}
\label{BBB}
In this appendix we will derive
explicit integral representations
of the scalar and
pseudoscalar current densities whose formal expressions are given in subsection
\ref{WChPT2pt}. Since we are dealing with the two-flavor case we can describe the
group manifold using the familiar parameterization of $SU(2)$
\begin{equation}U_0 = ( \cos{\alpha} + i \hat{n} \cdot \sigma \sin{\alpha}
)\ ,\end{equation}
\noindent where $\hat{n}$ is a three-dimensional unit vector, the $\sigma$'s are the
Pauli matrices and $0<\alpha< 2 \pi$. With this parameterization for an arbitrary
element of
$SU(2)$, the normalized group measure is
\begin{equation}\int dU_0 = \frac{1}{2\pi^2} \int d\Omega_{\hat{n}}
\int_0^{2\pi} d\alpha \sin{(\alpha)}^2.\end{equation}
\noindent
The partition function can thus be written in a more manageable way, using
this parameterization, as
\bea
\mathcal{Z}_{\rm NLO} &=&\frac{\mathcal{C}'}{2 \pi^2} \int_0^{2 \pi} d\alpha\,  \sin{(\alpha)}^2\, \text{exp}\left[ 2 m \Sigma^{\text{eff}} V
\cos{(\alpha)}
  - 16 a^2\, c_{2}^{\text{eff}}\,V\, \cos{(\alpha)}^2 \right] ,
\label{ZNLOnf2su2}
\eea
\noindent
and correspondingly for LO by dropping the superscript eff
and having a different normalization constant ${\cal C}$. As a check we obtain for $c_2=0$ the known result for the equal mass $SU(2)$ partition function $\mathcal{Z}_{\rm LO}= \mathcal{C} I_1(2\hm)/(2 \hm \pi)$ in terms of a modified Bessel function.
\newline
The expressions for the two-point scalar and pseudoscalar current
correlators derived in subsection \ref{WChPT2pt} can be rewritten as
\begin{eqnarray}\langle S_0(x) S_0(0)\rangle &=&
\frac{\left(\Sigma^{\text{eff}}\right)^2}{\mathcal{Z}_{\rm NLO}} \frac{\mathcal{C}'}{
2\pi^2} \int_0^{2 \pi} d\alpha\,  \sin{(\alpha)}^2\cos{(\alpha)}^2
e^{2 \hm^{\text{eff}}\cos{(\alpha)}
  - 16 \ha^2 c_{2}^{\text{eff}} \cos{(\alpha)}^2}
  \nn \\
  &-&
\frac{4 \Sigma^2}{F^2} \frac{\mathcal{C} \Delta(x)}{2 \pi^2\, \mathcal{Z}_{\rm LO}} \int_0^{2
\pi} d\alpha\,  \sin{(\alpha)}^2\left( \cos{(\alpha)}^2- 1\right)
e^{2 \hm \cos{(\alpha)}
  - 16 \ha^2c_{2} \cos{(\alpha)}^2}
\nn\\
&-&
\frac{64 \ha \Sigma c_3}{\sqrt{V}} \frac{\mathcal{C}}{2 \pi^2 \, \mathcal{Z}_{\rm LO}} \int_0^{2 \pi}
d\alpha\,  \sin{(\alpha)}^2\cos{(\alpha)}^3\,
e^{2 \hm \cos{(\alpha)}
  - 16 \ha^2c_{2} \cos{(\alpha)}^2 },
\end{eqnarray}
\begin{eqnarray}\langle P_b(x) P_b(0)\rangle &=& -
\frac{(\Sigma^{\text{eff}})^2\mathcal{C}'}{2 \pi^2\,
\mathcal{Z}_{\rm NLO}} \int_0^{2 \pi} d\alpha\,
\sin{(\alpha)}^2\left( \cos{(\alpha)}^2- 1\right)
e^{2 \hm^{\text{eff}} \cos{(\alpha)}
  - 16 \ha^2c_{2}^{\text{eff}} \cos{(\alpha)}^2}
  \nn \\
&+&
\frac{\Sigma^2}{F^2} \frac{\mathcal{C}\Delta(x)}{2 \pi^2\, \mathcal{Z}_{\rm LO}}
\int_0^{2 \pi} d\alpha\,  \sin{(\alpha)}^2\left( \cos{(\alpha)}^2+
2\right)
e^{2 \hm \cos{(\alpha)}
  - 16 \ha^2c_{2}\cos{(\alpha)}^2}
\nn \\
&+& \frac{ 16\ha c_3 \Sigma\ \mathcal{C}}{\sqrt{V} \pi^2\, \mathcal{Z}_{\rm LO}} \int_0^{2 \pi}
d\alpha\,  \sin{(\alpha)}^2\left( \cos{(\alpha)}^3-
\cos{(\alpha)}\right)
e^{2 \hm \cos{(\alpha)}
  - 16 \ha^2c_{2}\cos{(\alpha)}^2 }.
  \nn \\
\end{eqnarray}
\noindent
\normalsize

\section{Staggered Chiral Perturbation Theory for General $N_f$}
\label{D-SChPT}

In the following we report explicitly all the terms arising in the $\epsilon$-expansion up to order $\mathcal{O}(\epsilon^2)$ of the partition function $\mathcal{Z}_{\xi}(U_0)$ defined in the SChPT subsection \ref{SChPTnlo}. At LO $\mathcal{O}(\epsilon^0)$ one obtains
\small
\begin{center}
\begin{eqnarray}S^{(0)} &=& + \frac{1}{4}\int d^4x\, \text{Tr}\left[\partial_{\mu} \xi \partial_{\mu} \xi\right] - \frac{\Sigma V }{4} \text{Tr}\left[ M^{\dagger} U_0 + U_0^{\dagger} M\right] - a^2 V C_1 \text{Tr}\left(\gamma_5 U_0 \gamma_5 U_0^{\dagger\,}\right)\nn \\
  && - a^2 \frac{V C_3 }{2} \sum_{\mu}\, \left[ \text{Tr}\left( U_0 \gamma_{\mu} U_0 \gamma_{\mu}\right) + h.c.\right] - a^2 \frac{V C_4 }{2}\, \sum_{\mu} \left[  \text{Tr}\left( U_0 \gamma_{\mu 5} U_0 \gamma_{\mu 5}\right) + h.c.\right]\nn \\
  && - a^2 \frac{C_{2V}}{4}\, \sum_{\mu} \left[  \text{Tr}\left( U_0 \gamma_{\mu}\right) \text{Tr}\left( U_0 \gamma_{\mu }\right) + h.c. \right] - a^2 \frac{V C_{2A}}{4}\, \sum_{\mu} \left[  \text{Tr}\left( U_0 \gamma_{\mu 5}\right) \text{Tr}\left( U_0 \gamma_{\mu 5 }\right) + h.c.\right]\nn \\
  && - a^2 \frac{V C_{5V}}{4}\, \sum_{\mu} \left[  \text{Tr}\left( U_0 \gamma_{\mu}\right) \text{Tr}\left( U_0^{\dagger} \gamma_{\mu }\right) \right]- a^2 \frac{V C_{5A}}{4}\, \sum_{\mu} \left[  \text{Tr}\left( U_0 \gamma_{\mu 5}\right) \text{Tr}\left( U_0^{\dagger} \gamma_{\mu 5 }\right) \right] \nn \\
  && - a^2 V C_6\, \sum_{\mu < \nu} \text{Tr}\left[ U_0 \gamma_{\mu \nu } U_0^{\dagger} \gamma_{\mu \nu }\right],
\end{eqnarray}
 \end{center}
\noindent
\normalsize
while for the first order $O(\epsilon)$ we have $ S^{(1)} = 0\ ,$ and for the second order one finds
\small
%
\begin{center}
$ S^{(2)}=  \frac{1}{24 F^2} \int d^4x \text{Tr}\left[ [\partial_{\mu} \xi, \xi][\partial_{\mu} \xi, \xi]\right] +     \frac{\Sigma}{4 F^2}\int d^4x \text{Tr}\left[ M^{\dagger} U_0 \xi^2 + \xi^2 U_0^{\dagger} M\right] $\\
\vspace{0.2cm}$\, \, \, \,
+\int d^4x\Big[-  2 a^2 \frac{C_1}{F^2} \text{Tr}\left(\gamma_5 U_0 \xi \gamma_5 \xi \, U_0^{\dagger\,}  \right) +   a^2 \frac{C_1}{ F^2} \text{Tr}\left(\gamma_5 U_0 \xi^2 \gamma_5 U_0^{\dagger\,} \right) +  a^2 \frac{C_1}{F^2} \text{Tr}\left(\gamma_5 U_0 \gamma_5 \xi^2 U_0^{\dagger\,} \right)$ \\
\vspace{0.2cm}$+  a^2 \frac{C_3}{ F^2}\, \sum_{\mu} \left[ \text{Tr}\left( U_0 \xi \gamma_{\mu} U_0 \xi \gamma_{\mu} \right) + h.c. \right] +  a^2 \frac{C_3}{ F^2}\, \sum_{\mu} \left[ \text{Tr}\left( U_0 \xi^2 \gamma_{\mu} U_0  \gamma_{\mu}\right) + h.c. \right] $ \\
\vspace{0.3cm} $+  a^2 \frac{C_4}{ F^2}\, \sum_{\mu}\left[  \text{Tr}\left( U_0 \xi \gamma_{\mu 5}  U_0 \xi \gamma_{\mu 5}\right) + h.c.\right] + a^2 \frac{C_4}{ F^2}\, \sum_{\mu}\left[  \text{Tr}\left( U_0 \xi^2 \gamma_{\mu 5} U_0 \gamma_{\mu 5}\right) + h.c.\right]$\\
\vspace{0.3cm}$ +  a^2 \frac{C_{2V}}{2 F^2}\, \sum_{\mu} \left[  \text{Tr}\left( U_0 \xi \gamma_{\mu }\right)\text{Tr}\left( U_0 \xi \gamma_{\mu}\right) + h.c.\right] +   a^2 \frac{C_{2V}}{2 F^2}\, \sum_{\mu} \left[  \text{Tr}\left( U_0 \xi^2 \gamma_{\mu }\right) \text{Tr}\left( U_0 \gamma_{\mu}\right) + h.c.\right]$\\
\vspace{0.3cm}$+ a^2 \frac{C_{2A}}{2 F^2}\, \sum_{\mu} \left[  \text{Tr}\left( U_0 \xi \gamma_{\mu 5}\right)\text{Tr}\left( U_0 \xi \gamma_{\mu 5}\right) + h.c.\right] +   a^2 \frac{C_{2A}}{2 F^2}\, \sum_{\mu} \left[  \text{Tr}\left( U_0 \xi^2 \gamma_{\mu 5}\right) \text{Tr}\left( U_0 \gamma_{\mu 5}\right) + h.c. \right]$\\
\vspace{0.2cm}$- a^2 \frac{C_{5 V}}{2 F^2}\, \sum_{\mu} \left[  \text{Tr}\left( U_0 \xi \gamma_{\mu }\right) \text{Tr}\left( \xi U_0^{\dagger} \gamma_{\mu }\right) - \frac{1}{2} \text{Tr}\left( U_0 \xi^2  \gamma_{\mu}\right) \text{Tr}\left( U_0^{\dagger\,} \gamma_{\mu }\right) - \frac{1}{2} \text{Tr}\left( U_0 \gamma_{\mu}\right) \text{Tr}\left( \xi^2 U_0^{\dagger\,} \gamma_{\mu }\right)\right]$\\
 \vspace{0.2cm}$- a^2 \frac{C_{5 A}}{2 F^2}\, \sum_{\mu} \left[  \text{Tr}\left( U_0 \xi \gamma_{\mu 5}\right) \text{Tr}\left( \xi U_0^{\dagger} \gamma_{\mu 5 }\right) - \frac{1}{2} \text{Tr}\left( U_0 \xi^2  \gamma_{\mu 5}\right) \text{Tr}\left( U_0^{\dagger\,} \gamma_{\mu 5}\right) - \frac{1}{2} \text{Tr}\left( U_0 \gamma_{\mu 5}\right) \text{Tr}\left( \xi^2 U_0^{\dagger\,} \gamma_{\mu 5 }\right)\right]$\\
 \vspace{0.2cm}$- 2 a^2 \frac{C_6}{F^2}\, \sum_{\mu < \nu} \text{Tr}\left( U_0 \xi \gamma_{\mu \nu } \xi U_0^{\dagger} \gamma_{\mu \nu }\right) +  a^2 \frac{C_6}{F^2}\, \sum_{\mu < \nu}\left[ \text{Tr}\left( U_0 \xi^2 \gamma_{\mu \nu } U_0^{\dagger} \gamma_{\mu \nu }\right) + \text{Tr}\left( U_0 \gamma_{\mu \nu } \xi^2 U_0^{\dagger} \gamma_{\mu \nu }\right)\right] \Big].$\\
\end{center}
\noindent
\normalsize
Now one can perform the Gaussian integrals over the fluctuations, and one finds that
%
\begin{center}
${\cal Z}_{\xi(U_0)} = \mathcal{N}\left( 1 - \frac{m V \Sigma}{4 F^2} \frac{16 N_f^2-1}{4 N_f} \Delta(0)\, \text{Tr}\left[ U_0 + U_0^{\dagger} \right] - \frac{ 8 a^2 C_1 V}{F^2}\, N_f \Delta(0) \text{Tr}\left[ U_0 \gamma_5 U_0^{\dagger\, }\gamma_5 \right]  \right.$\\
 \vspace{0.2cm}$\left.- \frac{  a^2 [C_3(16 N_f^2 -2)+ 2\, C_{2V} N_f]}{4 N_f\, F^2}\, \Delta(0) V \sum_{\mu} \left[\text{Tr}\left( U_0 \gamma_{\mu} U_0 \gamma_{\mu} \right) + h.c. \right]  \right.$\\
\vspace{0.2cm}$ \left.- \frac{  a^2 [C_4(16 N_f^2 -2)+ 2\, C_{2A} N_f]}{4 N_f\, F^2}\, \Delta(0) V \sum_{\mu} \left[\text{Tr}\left( U_0 \gamma_{\mu5} U_0 \gamma_{5\mu} \right) + h.c.\right]  \right.$\\
\vspace{0.2cm}$  \left.- \frac{  a^2 [C_{2V}(16 N_f^2 -2)+ 8 C_{3} N_f]}{8 N_f\, F^2}\, \Delta(0) V \sum_{\mu} \left[\text{Tr}\left( U_0 \gamma_{\mu}\right) \text{Tr}\left(U_0 \gamma_{\mu} \right) + h.c. \right] \right.$\\
\vspace{0.2cm}$\left.- \frac{ a^2 [C_{2A}(16 N_f^2 -2)+ 8 C_{4} N_f]}{8 N_f\, F^2}\, \Delta(0) V \sum_{\mu} \left[ \text{Tr}\left( U_0 \gamma_{\mu5}\right) \text{Tr}\left(U_0 \gamma_{5 \mu} \right) + h.c. \right]\right.$\\
\vspace{0.2cm}$ \left.- \frac{4 a^2 C_{5 V}  N_f}{2 \, F^2}\, \Delta(0) V \sum_{\mu} \text{Tr}\left( U_0 \gamma_{\mu}\right) \text{Tr}\left(U_0^{\dagger} \gamma_{\mu} \right)\right.$\\\vspace{0.2cm}$\left. - \frac{ 4 a^2 C_{5 A}  N_f}{2 \, F^2}\, \Delta(0) V \sum_{\mu} \text{Tr}\left( U_0 \gamma_{\mu5}\right) \text{Tr}\left(U_0^{\dagger} \gamma_{5 \mu} \right) \right.$\\
\vspace{0.2cm}$  \left.- \frac{ 8 a^2 C_6 N_f}{\, F^2}\, \Delta(0) V\sum_{\mu < \nu} Tr\left[ U_0 \gamma_{\mu \nu } U_0^{\dagger} \gamma_{\mu \nu }\right]\right).$
\end{center}
%
Because in this appendix we didn't have to use any group integral identities the same relations hold for fixed index, by adding $\det[U_0^\nu]$ inside the zero-mode group integral.


\end{document}